\documentclass[12pt,a4paper]{article}
\usepackage[utf8]{inputenc}
\usepackage{graphicx} 
\usepackage{authblk}
\usepackage{float}
\usepackage{gensymb}
\usepackage{comment}
\usepackage{hyperref}
\usepackage{ragged2e} 
\usepackage[dvipsnames]{xcolor} 
\usepackage{amssymb}
\usepackage{multirow}
\usepackage[labelfont=bf]{caption}
\usepackage[labelsep=period]{caption}

\usepackage{setspace}
\usepackage{lineno}

\usepackage[margin=2.0cm]{geometry}
 
\usepackage[superscript,biblabel]{cite}

\title{\vspace{-2cm}\LARGE Electroferrofluids with Non-Equilibrium Voltage-Controlled Magnetism, Interfaces, and Patterns}

\author[1]{\normalsize Tomy Cherian$^\dagger$}
\author[1]{Fereshteh Sohrabi$^\dagger$}
\author[1]{Carlo Rigoni}
\author[1]{Olli Ikkala}
\author[1]{Jaakko V. I. Timonen*}
\affil[1]{\small Department of Applied Physics, Aalto University School of Science, Puumiehenkuja 2, 02150 Espoo, Finland}

\date{}
\begin{document}
\maketitle

\noindent\textbf{Materials with continuous dissipation can exhibit responses and functionalities that are not possible in thermodynamic equilibrium.\cite{cross2009pattern,grzybowski2017dynamic,sorrenti2017non} While this concept is well-known, a major challenge has been the implementation: how to rationally design materials with functional non-equilibrium states and quantify the dissipation? Here we address these questions for the widely used colloidal nanoparticles \cite{yin2005colloidal} that convey several functionalities.\cite{grzelczak2019stimuli} We propose that useful non-equilibrium states can be realised by creating and maintaining steady-state nanoparticle concentration gradients by continuous injection and dissipation of energy. We experimentally demonstrate this with superparamagnetic iron oxide nanoparticles that in thermodynamic equilibrium form a homogeneous functional fluid with a strong magnetic response (a ferrofluid).\cite{zhang2019flexible} To create non-equilibrium functionalities, we charge the nanoparticles with anionic charge control agents to create electroferrofluids where nanoparticles act as charge carriers that can be driven with electric fields and current to non-homogeneous dissipative steady-states. The dissipative steady-states exhibit voltage-controlled magnetic properties and emergent diffuse interfaces. The diffuse interfaces respond strongly to external magnetic fields, leading to dissipative patterns that are not possible in the equilibrium state. We identify the closest non-dissipative analogues of these dissipative patterns, discuss the differences, and highlight how pattern formation in electroferrofluids is linked to dissipation that can be directly quantified. Beyond electrically controlled ferrofluids and patterns, we foresee that the concept can be generalized to other functional nanoparticles\cite{grzelczak2019stimuli} to create various scientifically and technologically relevant non-equilibrium states with optical, electrical, catalytic, and mechanical responses that are not possible in thermodynamic equilibrium.}

\vspace{10mm}
\justify
Dissipative microscopic structures that exist far from thermodynamic equilibrium through sustained energy dissipation can exhibit responses and functionalities unparalled with systems in thermodynamic equilibrium.\cite{whitesides2002self,cross2009pattern}  While many conceptually important macroscopic model systems have been developed where the role of dissipation in assembly and functionalities can be directly visualized,\cite{grzybowski2000dynamic,timonen2013switchable} realizations of microscopic systems with non-equilibrium responses have been scarce. Only very recently synthetic molecular systems driven by chemical fuels\cite{boekhoven2015transient,maiti2016dissipative} 
and colloidal systems of nanoparticles (NPs) driven by UV light\cite{klajn2009writing}, oscillating pH \cite{lagzi2010nanoparticle}, chemical fuels \cite{sawczyk2017out,grotsch2018dissipative} or electric field\cite{yu2017reversible} have started to emerge with promising non-equilibrium responses. However, there is still need for novel systematic approaches to drive non-equilibrium steady-states with emergent functionalities that can be maintained with reasonable and practical energy consumption.

\begin{figure}
\centering
\includegraphics{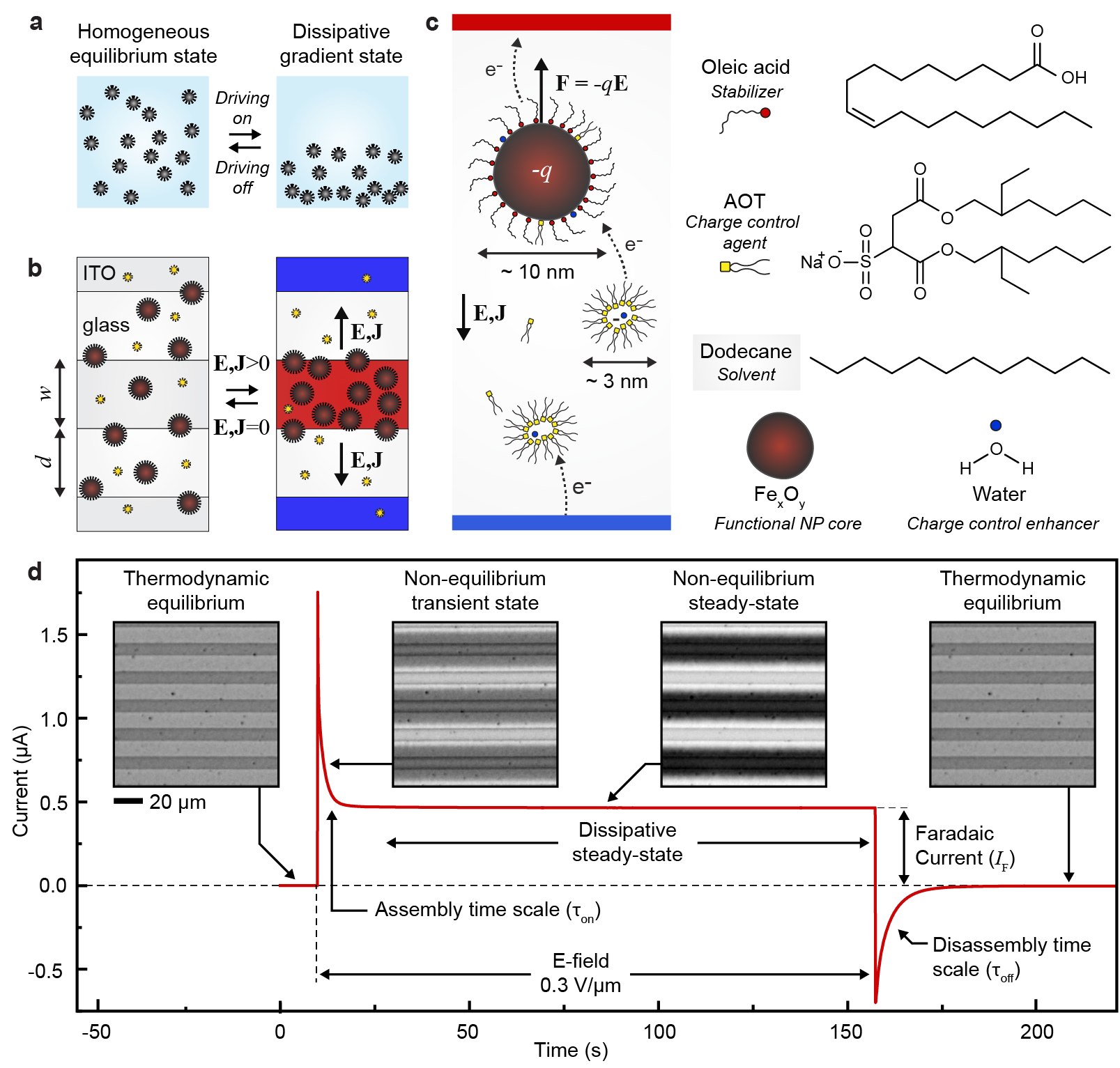}
\caption{
\textbf{Driving non-equilibrium steady-states in colloidal NP dispersions with an electric field.} 
\textbf{a,} A conceptual scheme of driving a NP dispersion out of the homogeneous equilibrium state to create a non-equilibrium steady-state with sustained gradients in NP concentration.
\textbf{b,} A scheme of a practical realization of the concept by driving weakly charged NPs in non-polar solvent with electric field using planar microelectrodes.
\textbf{c,} A scheme showing the composition of the electroferrofluid and the expected charge transfer processes.
\textbf{d,} Electric current as a function of time across the electroferrofluid (6 \% iron oxide NPs and 150 mM AOT in dodecane) in a microelectrode cell ($U$ = 3 V, $E$ = 0.3 V/$\mu$m). Insets show microscopy images of the cell at different points of time, revealing formation of steady-state concentration gradients during the dissipation and their decay after dissipation is turned off (Supplementary Video S1).}
\label{Fig1}
\end{figure}

In this article, we propose that novel responses and functionalities can emerge when homogeneous nanoparticle dispersions are driven out of the thermodynamic equilibrium into inhomogeneous steady-states that are sustained by continuous dissipation (Fig. 1a). Because the responses of colloidal dispersions (optical, mechanical, magnetic, and so on) depend on NP concentration, these responses become dissipatively controlled. In principle, the concept (Fig. 1a) can be realized with various driving and dissipation mechanisms. Here we chose electrical driving (Fig. 1b) and give demonstrations using superparamagnetic iron oxide nanoparticles stabilized with oleic acid in dodecane (Fig. 1c, core diameter $d=8.3 \pm 2.6$ nm, see Methods and Extended Data Fig. 1 for synthesis and characterization). When dispersed in a non-polar solvent, these NPs form stable sterically stabilized and magnetically responsive homogeneous dispersion known as a ferrofluid with diverse technological applications and scientific interest.\cite{zhang2019flexible} To make them electrically responsive, we add bis(2-ethylhexyl) sulfosuccinate sodium salt (Aerosol OT, AOT) that acts as a charge control agent\cite{kemp2010nanoparticle,martin2010charged, shevchenko2006structural,yu2017reversible,bulavchenko2008electrophoretic} forming charge-carrying reverse micelles\cite{karvar2015charging} and charging the NPs (Fig. 1c).\cite{bulavchenko2008electrophoretic}. We further incubate the mixture in a humidity chamber to control the amount of water that acts as charge control enhancer, to yield a complex electrically active mixture.

We drive the homogeneous mixture out of thermodynamic equilibrium by applying an electric field using microelectrode cells with planar interdigitated indium tin oxide (ITO) electrodes (width \textit{w} = 10 $\mathrm{\mu m}$ and spacing \textit{d} = 10 $\mathrm{\mu m}$, cell height \textit{h} = 4 $\mathrm{\mu m}$, see the inset of Fig. 2a). The electric field is approximately uniform and in-plane between the microelectrodes with an average value of $E=U/d$ (Extended Data Fig. 2). Application of the electric field results in a current peak that stabilizes in a few seconds to a steady-state value $I_{\mathrm{F}}$ (Fig. 1d). At microscopic level, this is accompanied by redistribution of the NPs from the initially homogeneous state towards the vicinity of the positive electrodes, suggesting that the NPs are negatively charged (Supplementary Video S1). The NP concentration gradients are sustained as long as the system is being driven by the electric field \textbf{E} and the resulting current \textbf{J}. Once the field is switched off, the current relaxes to zero and the NPs redisperse to form again the equilibrium homogeneous state. It is important to note that these non-equilibriums states can be sustained almost indefinitely (in contrast to often seen transient behaviors \cite{lin2014pattern,sawczyk2017out,grotsch2018dissipative}). 

The role of the non-polar environment and the charge control agent (AOT) is important for the electroferrofluid behavior. Control experiments carried out in polar environment (aqueous dispersion of highly charged iron oxide nanoparticles) show no gradient formation - instead electrodeposition of the NPs on the ITO electrodes takes place (Extended Data Fig. 3). In a non-polar environment, the behavior is highly dependent on the concentration of AOT as is seen both in electrical conductivity (Fig. 2a) and microscopic behavior of the NPs (Fig. 2b). At small or vanishing AOT concentrations, the (electro)ferrofluid does not conduct nor dissipate energy (Fig. 2a) and the NPs aggregate irreversibly under applied electric field (Fig. 2b, Supplementary Video S2). The poor conductivity is in agreement with lack of charge carriers (including both charged reverse micelles and charged NPs). The NP aggregation is attributed to generation of positively and negatively charged NPs near the positive and negative electrodes, respectively, and their Coulombic attraction (Fig. 2c). Because of the non-polar environment with low relative permittivity $\epsilon_{\mathrm{r}} \approx$ 2, even two opposing elementary charges $e$ localized at centers of two NPs exhibit Coulombic interaction energy that overcomes the thermal energy $k_{\mathrm{B}} T$ as $E_{\mathrm{C}} =-e^2/(4\pi\epsilon_{\mathrm{r}}\epsilon_0 d) \approx 3 k_{\mathrm{B}} T$ ($d$ = 10 nm). In contrast, at higher AOT concentration near 50 mM and above, the NP aggregation is suppressed and gradients appear instead (Fig. 2b), and intensify further as AOT concentration is increased (Extended Data Fig. 4). This is accompanied by increase in conductivity and dissipation (Fig. 2a). Because an electroferrofluid containing 150 mM AOT is significantly more conductive than plain 150 mM AOT in dodecane (with reverse micelles as the only charge carriers), it appears that there is strong interaction between AOT and NPs that leads to significant charging of the NPs that act as additional charge carriers (Fig. 2a). Together with the spatial accumulation of nanoparticles near the positive electrodes, these observations suggest a mechanistic picture wherein negative charge is injected to reverse micelles near the negative electrode, which then further migrate electrophoretically towards the positive electrode where they meet the cloud of NPs, and finally donate the charge to the NPs that carry it ultimately to the positive electrode (Fig. 1c).

\begin{figure}
\centering
\includegraphics{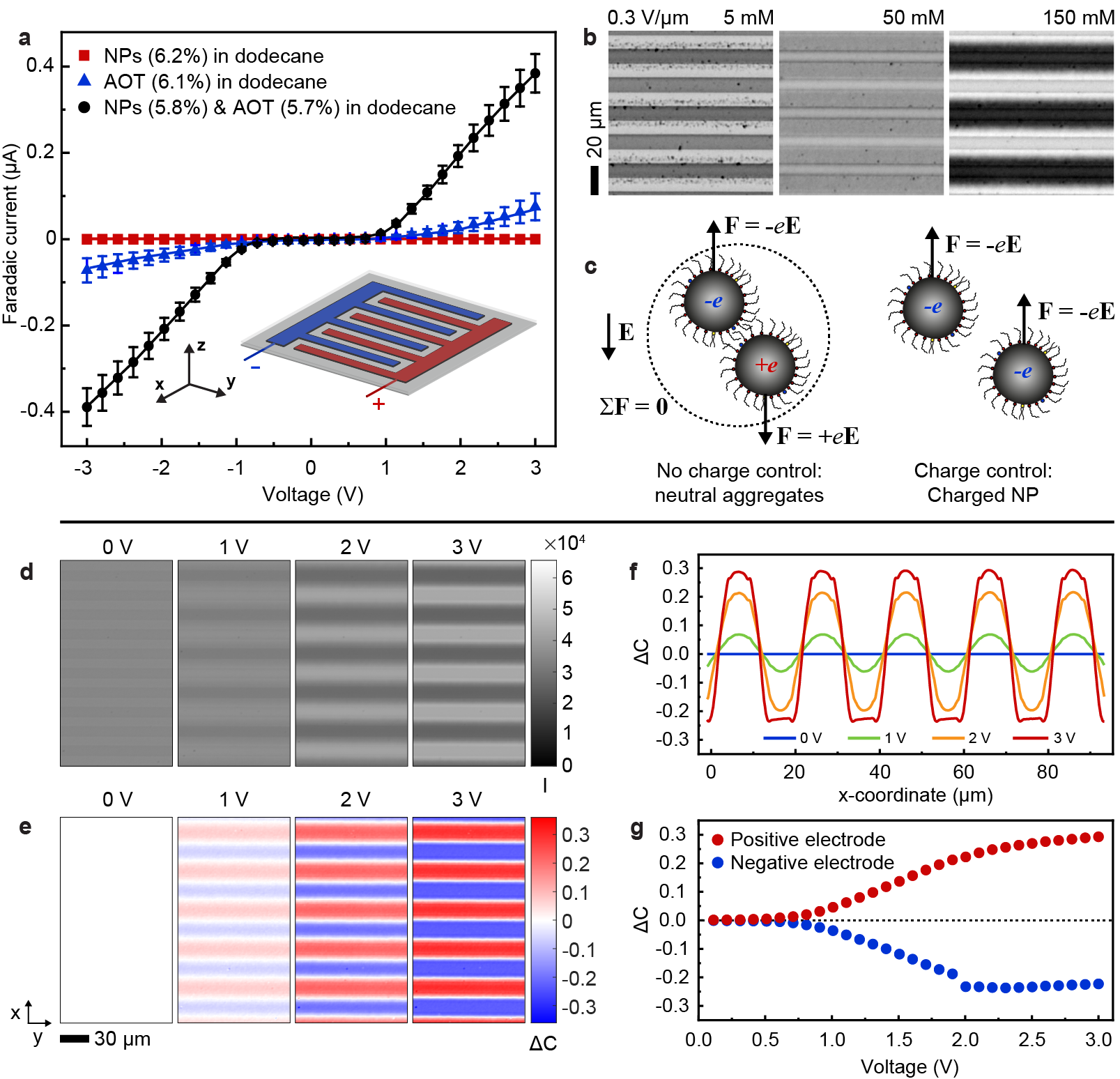}
\caption{
\textbf{Magnetic NPs carrying similar charges enable electrically controlled magnetism in electroferrofluids.} 
\textbf{a,}  "IV-curves" of NPs in dodecane, AOT in dodecane, and both NPs and AOT in dodecane. All percentages in the legend are volume fractions (6.1\% AOT corresponds to 150 mM). Error bars indicate standard deviations from 3 independent measurements carried out in separete microelectrode cells.
\textbf{b,} Optical microscopy images showing the steady-state of NPs at different AOT concentrations at a fixed voltage of $U$ = 3 V.
\textbf{c,} Scheme of NP charging state and interactions at low (left) and high (right) AOT concentrations.
\textbf{d,} Unprocessed optical microscopy images (16-bit gray scale) showing the steady-state NP gradients at different voltages in dodecane with 150 mM AOT.
\textbf{e,} Colormaps showing $\Delta C$ calculated using Eq. (1) from the images shown in \textbf{d}. 
\textbf{f,} One dimensional plots of $\Delta C$ as a function of x-coordinate obtained from the colormaps in \textbf{e}.
\textbf{g,} $\Delta C$ as a function of applied voltage on top of the negative and positive electrodes, showing accumulation of NPs at the positive electrode.
}
\label{Fig2}
\end{figure}

The electrically controlled dissipative build-up of concentration gradients of superparamagnetic NPs allows electrically controlled non-equilibrium magnetic responses. In the first approximation, the magnetic susceptibility $\chi$ and the saturation magnetization $M_{\mathrm{S}}$ of a NP dispersion is linearly proportional to the NP concentration $c$. As in voltage-controlled magnetism in solid state materials,\cite{ohno2000electric-field,weisheit2007electric,bauer2014magneto-ionic} it is difficult to perform direct local magnetic measurements on the voltage-controlled state, and we thus rely on approximate quantification by analysing optical transmittance changes through the sample that reflect on the NP concentration. Approximately, by the Beer-Lambert law, light intensity transmitted through the microelectrode cell in any location depends on the local concentration of NPs as $I\propto e^{-\alpha h c}$ where $\alpha$ is extinction coefficient of the NPs. It follows that we can define for each spatial location (pixel) a normalized change in NP concentration ($\Delta C$) that is proportional to the difference of the non-equilibrium NP concentration ($c$) and the equilibrium concentration ($c_0$), and at the same time can be calculated from the ratio between the transmitted light intensity in the non-equilibrium state ($I$) and in the equilibrium state ($I_0$) as
\begin{equation}
\Delta C = \alpha h (c-c_0) = - \ln (I/I_0) . 
\end{equation}
For non-interacting NPs, $\chi \propto c$ and $M_{\mathrm{S}} \propto c$, and thus $\Delta\chi$ = $\chi-\chi_0\propto \Delta C$ and $\Delta M_{\mathrm{S}}$ = $M_{\mathrm{S}}-M_{\mathrm{S0}} \propto \Delta C$, so that $\Delta C$ is a fingerprint parameter linearly proportional to the local change in both the susceptiblity and the saturation magnetization. 

With help of Eq. (1), the unprocessed microscopy images of the steady-state gradients (Fig. 2d) can be converted to maps of $\Delta C$ (Fig. 2e, see Extended Data Fig. 5 for details). The magnetic response increases near the positive electrodes and conversely decreases near the negative electrodes with the increasing voltage (Fig. 2e). Profiles of $\Delta C$ perpendicular to the electrodes show that the magnetic responses vary smoothly, approximately sinusoidally, at small electric fields (Fig. 2f, $U$ =  1 V and $U$ = 2 V). At higher fields, the smooth perturbations are not anymore possible because of complete depletion of NPs from the vicinity of the negative electrodes, leading to flattening of $\Delta C$ (Fig. 2f, $U$ = 3 V). The build-up of magnetic contrast increases non-linearly with the applied voltage near both positive and negative electrodes (Fig. 2g). At low voltages, below 0.5 V, $\Delta C \approx $ 0, resembling the dependence of the electric current on voltage (Fig. 2a). This suggests that current and dissipation are indeed critical for the formation of NP gradients and the resulting voltage-controlled magnetism. 

The key feature of ferrofluids is their response to magnetic fields that enables several functionalities.\cite{zhang2019flexible} With electoferrofluids, all these functionalities become also voltage-controlled. Here we consider the elementary response of the electroferrofluid in the non-equilibrium gradient state to uniform magnetic field applied either perpendicular (Fig. 3) or parallel (Fig. 4) to the microelectrode cell (see Extended Data Fig. 6 for details of the experimental setup). In the perpendicular case (Fig. 3a), the steady-state NP gradients remain unchanged in weak magnetic fields, but become unstable above a threshold magnetic field strength ($B_c$) of few mT (Fig. 3b). The instability is seen directly as microscopic fluctuations in the NP gradient that after few seconds stabilize to a well-defined one-dimensional steady-state pattern (Fig. 3b, Supplementary Video S3). The pattern wavelength $\lambda$ is close to the spacing between electrodes and decreases with increasing magnetic field strength (Fig. 3c,d, Supplementary Video S4). Similarly, increasing electric field strength (and thus susceptibility and magnetization contrast, Fig. 2) leads to decrease in the pattern periodicity (Fig. 3d). Kymographic analysis shows that the patterns are not entirely static over time but experience varying degrees of fluctuations depending on the magnetic field strength (Fig. 3e). Especially near the threshold ($B \approx B_{\mathrm{c}}$), i.e. transition from a steady gradient state (Fig. 3e, 3.2 mT) to a steady pattern (Fig. 3e, 4.8 mT), there exists an intermediate fluctuating state wherein the pattern appears and disappears randomly in different locations (Fig. 3e, 4.4 mT). Thus, overall, the behavior of the electroferrofluid in the combination of perpendicular magnetic field and in-plane electric field forms a two-dimensional phase diagram with one homogeneous equilibrium (EQ) state and three non-equilibrium (NE1-NE3) states: stable gradient state, steady pattern state, and a narrow segment corresponding to the fluctuating pattern state in-between (Fig. 3f).

\begin{figure}
\centering
\includegraphics{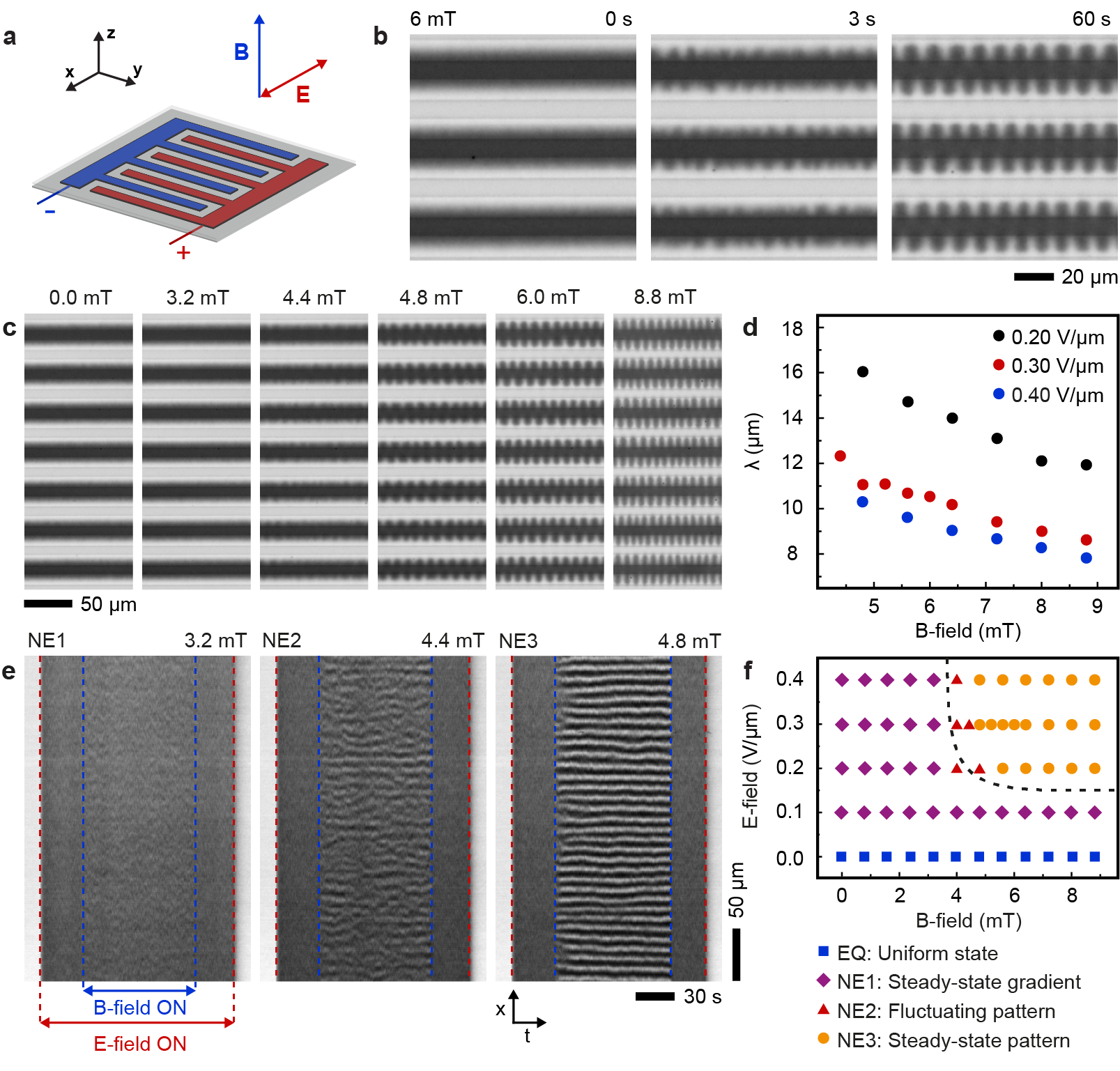}
\caption{
\textbf{Response of electroferrofluid (6 \% iron oxide NPs and 150 mM AOT) to uniform perpendicular magnetic field $\textbf{B}$ = $B\hat{\textbf{z}}$.} 
\textbf{a,} Scheme of the microelectrode cell with directions of electric and magnetic fields indicated.
\textbf{b,} Microscopy image series showing the response of a NP gradient state (0.30 V/${\mathrm{\mu m}}$) to 6 mT magnetic field (Video S3).
\textbf{c,} Microscopy images of the steady-state with electric field (0.30 V/${\mathrm{\mu m}}$) and magnetic field (0 mT to 8.8 mT) applied (Video S4). 
\textbf{d,} Average pattern periodicity ($\lambda$) as a function of magnetic field strength for three different electric field strengths (see Extended data Fig. 8 for corresponding microscopy images).
\textbf{e,} Kymographs obtained at three magnetic field strengths and fixed electric field strength (0.30 V/${\mathrm{\mu m}}$) near the threshold line.
\textbf{f,} Phase diagram showing the four states of the electroferrofluid as a function of in-plane electric field and perpendicular magnetic field. Approximate location of the threshold for instability and pattern formation is indicated with the dashed line. The phase diagram was constructed based on steady-state snapshots taken 1 min after turning on the magnetic field.
}
\label{Fig3}
\end{figure}

The behavior of the electroferrofluid in magnetic field parallel to the microelectrode cell and perpendicular to the microelectrodes (Fig. 4a) is qualitatively similar as in perpendicular magnetic field (Fig. 3) with some exceptions. Similar to the perpendicular case, the gradient state becomes unstable above a threshold magnetic field, but the appearance of the patterns is different and formation time scales longer (Fig. 4b, Video S5). The pattern periodicity is affected by magnetic field strength (Fig. 4c, Video S6), but the dependency is non-monotonic (Fig. 4d, Extended data Fig. 9). We attribute this complexity to larger amplitude of the pattern, leading to strong interactions of the patterns between neighbouring positive electrodes (Fig. 4c) that is weak or absent in the perpendicular case (Fig. 3c). Similarly to perpendicular case, there is one equilibrium state and three non-equilibrium states including the fluctuating states (Fig. 4e,f).

\begin{figure}
\centering
\includegraphics{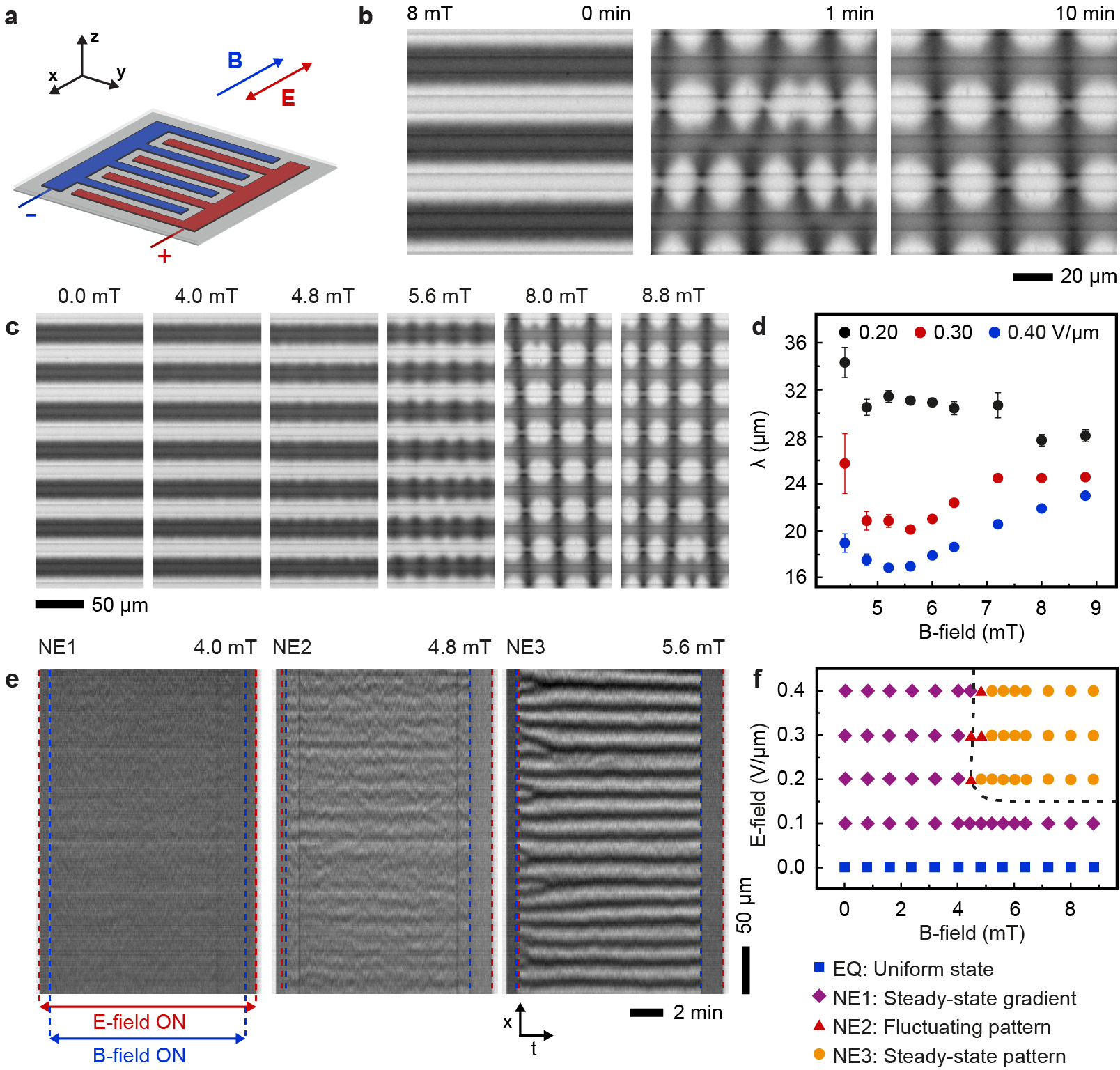}
\caption{
\textbf{Response of electroferrofluid (6 \% iron oxide NPs and 150 mM AOT) to uniform parallel magnetic field $\textbf{B}$ = $B\hat{\textbf{x}}$.} 
\textbf{a,} Scheme of the microelectrode cell with directions of electric and magnetic fields indicated.
\textbf{b,} Microscopy image series showing the response of a NP gradient state (0.30 V/${\mathrm{\mu m}}$) to 8 mT magnetic field (Video S5).
\textbf{c,} Microscopy images of the steady-state with electric field (0.30 V/${\mathrm{\mu m}}$) and magnetic field (0 mT to 8.8 mT) applied (Video S6). 
\textbf{d,} Average pattern periodicity ($\lambda$) as a function of magnetic field strength for three different electric field strengths (see Extended data Fig. 9 for corresponding microscopy images).
\textbf{e,} Kymographs obtained at three magnetic field strengths and fixed electric field strength (0.30 V/${\mathrm{\mu m}}$) near the threshold line.
\textbf{f,} Phase diagram showing the four states of the electroferrofluid as a function of in-plane electric field and parallel magnetic field. Approximate location of the threshold for instability and pattern formation is indicated with the dashed line. The phase diagram was constructed based on steady-state snapshots taken 1 min after turning on the magnetic field.
}
\label{Fig4}
\end{figure}

It is interesting to compare the magnetic response of the electroferrofluid (Fig. 3, 4) to the behavior of classic equilibrium ferrofluids under similar conditions. In particular, classic ferrofluids exhibit two famous and thoroughly-studied\cite{rosensweig2013ferrohydrodynamics,richter2005two,dickstein1993labyrinthine} equilibrium patterns that are known since 1960s: the Rosensweig pattern and the labyrinthine pattern (Fig. 5a).\cite{cowley1967interfacial, rosensweig1983labyrinthine} These classic patterns are driven by minimization of the total energy of the system (with contributions from magnetostatic energy, interfacial energy and gravitational energy). Analogously, the electroferrofluid patterns can be classified as \textit{labyrinthine-like} (Fig. 3) and \textit{Rosensweig-like} (Fig. 4) based on the orientation of the magnetic field with respect to the diffuse interface. However, here the similarities end and differences start. Firstly, the appearance and length scales of the patterns are different (Fig. 5a). Secondly, the interfacial tension plays a critical role in the classic patterns suppressing the pattern formation - but in electroferrofluids such interface does not exist since the interface is diffuse (Fig. 5b). Thirdly, while both classic equilibrium and the dissipative patterns exhibit a threshold for onset of pattern formation, only electroferrofluids exhibit the state of a fluctuating pattern. This is likely linked to the diffuse nature of the interface and is thus absent in the classic systems with thermodynamic, molecular scale phase boundaries (Fig. 5b). Fourthly, the electroferrofluid patterns are dissipative features that can exist only with continuous energy feed and dissipation, which can be directly quantified.

The power dissipated in electroferrofluids in any non-equilibrium steady-state is given simply as $P$ = $U \cdot I_{\mathrm{F}}$. Under typical experimental conditions, $U$ = 3 V and $I_\mathrm{F}\approx 400 $ nA, the dissipated power is very modest 1.2 ${\mathrm{\mu W}}$ for a microelectrode cell with a surface area of ca. 1 cm$^2$. The pattern formation in magnetic field (Fig. 3, 4) is associated with a measurable change in Faradaic current ($\Delta I$, Fig. 5c). This implies that dissipated power can be divided into two components as $P$ = $P_0 + \Delta P$, where $P_0$ is power dissipated in electric field only and $\Delta P$ = $U \cdot \Delta I$ is the increase due to the application of a magnetic field. Increase in current is seen only above the threshold magnetic field strengths for pattern formation (Fig. 5d, see Extended Data Fig. 10 for details on extraction of $\Delta I$ from $I(t)$ curves). Therefore, pattern formation is intimately linked to dissipation in the electroferrofluids. 

\begin{figure}
\centering
\includegraphics{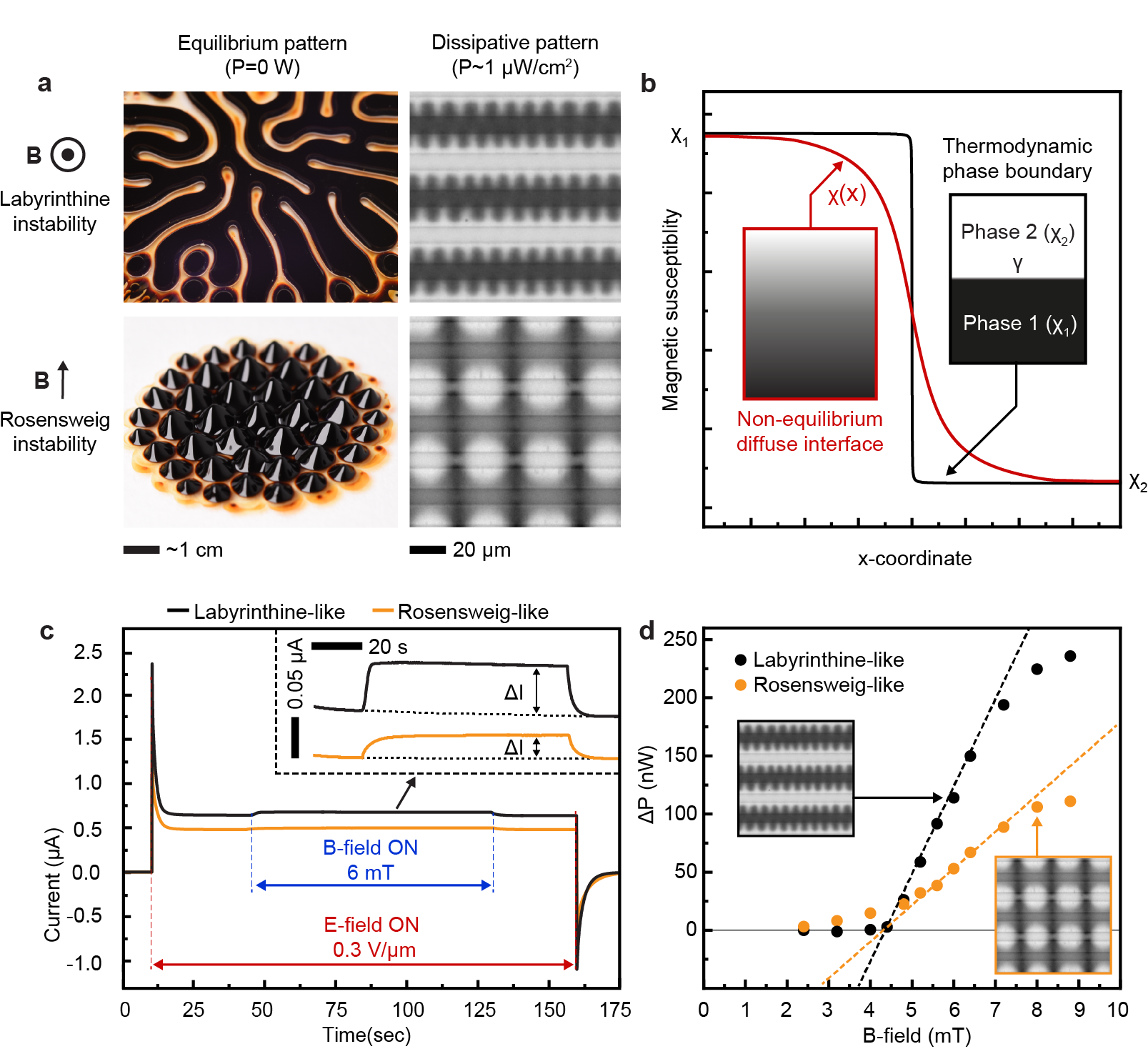}
\caption{\textbf{Equilibrium pattern analogies and quantification of dissipation in electroferrofluids.}
\textbf{a,} Images of the classical equilibrium Rosensweig and labyrinthine patterns and the analogous dissipative \textit{Rosensweig-like} and \textit{labyrinthine-like} patterns. Electroferrofluid was used in all the images. 
\textbf{b,} Scheme of the behavior of magnetic susceptibility near a classic equilibrium interface between two fluids with different magnetic susceptiblities and near a dissipative diffuse interface reported in this work.
\textbf{c,} Changes in the electric current upon the formation of Rosensweig-like and labyrinthine-like patterns (\textit{B} = 6 mT, \textit{E} = 0.30 V/${\mathrm{\mu m}}$).
\textbf{d,} Change in power dissipated as a function of applied magnetic field in the steady-sate labyrinthine-like and Rosensweig-like patterns at \textit{E} = 0.30 V/${\mathrm{\mu m}}$. 
}\label{Fig5}
\end{figure}

In conclusion, we have demonstrated a facile approach to drive colloidal NPs out of thermodynamic equilibrium with an electric field to create non-equilibrium dissipative states and responses. We foresee that this concept that we demonstrated using superparamagnetic nanoparticles will open new avenues in several  fields of research. Firstly, the  electroferrofluids with voltage-controlled magnetism (Fig. 2) can replace classic ferrofluids in many existing applications,\cite{zhang2019flexible} leading to a wide variety of functional materials and surfaces\cite{wang2018multifunctional} that can be switched not only with magnetic field  but also with electric field. Secondly, the concept of diffuse non-equilibrium interfaces replacing classic thermodynamic phase boundaries (Fig. 5b) can be extended, in principle, to any liquid-liquid system to create other dissipative analogues of the many classic interfacial responses and patterns.\cite{cross2009pattern} Thirdly, and perhaps most importantly, there is no foreseeable limitation to why other kinds of NPs with broadly tunable functionalities\cite{grzelczak2019stimuli} and interactions\cite{batista2015nonadditivity} enabled by the revolution in control of NP synthesis\cite{yin2005colloidal} could not be driven similarly out of thermodynamic equilibrium with electric or other fields to create non-equilibrium steady-states and diffuse interfaces. This will immediately lead to a generalised scheme for creating non-equilibrium plasmonic,\cite{pastoriza2018plasmonic} photonic,\cite{garcia2007photonic} mechanical,\cite{trappe2001jamming} and other responses that are not possible in thermodynamic equilibrium. The very modest Faradaic current and dissipation (a cm-scale microelectrode cell drawing 400 nA current out of two standard 1.5 V AA batteries could theoretically reside in the steady-state for over 500 years) suggests feasible technological applications as e.g. tunable optical devices or in mobile devices.

\clearpage
\justify


{\large \textbf{Methods} \par}

\justify \textbf{Materials}\\
Iron (III) chloride hexahydrate ($\mathrm{FeCl_3\cdot6H_2O}$, $\geq$ 99\%, Sigma-Aldrich), iron (II) sulfate heptahydrate ($\mathrm{FeSO_4\cdot7H_2O}$, $\geq$ 99\%, ACS reagent, Sigma-Aldrich), ammonium hydroxide ($\mathrm{NH_4OH}$, 28-30\%, ACS reagent, Sigma-Aldrich), oleic acid (90\% technical grade, Sigma-Aldrich), acetone ($\geq$ 99.8\%, Fisher Scientific), toluene ($\geq$ 99.7\%, ACS reagent, Sigma-Aldrich), n-dodecane (99\%, anhydrous, Acros Organics), docusate sodium salt (AOT, $\geq$99\%, anhydrous, Sigma-Aldrich), magnesium nitrate hexahydrate, ($\mathrm{Mg(NO_3)_2\cdot6H_2O}$, 99\%, ACS reagent, Sigma Aldrich), and microelectrode cells with interdigitated ITO electrodes (Instec IPS10X10A040uNOPI) were used as received.

\justify \textbf{Synthesis of stock dispersion of iron oxide NPs in toluene}\\
Iron oxide NPs were synthesised by coprecipitation method\cite{Massart1981preparation}, and stabilized with oleic acid  in toluene. Briefly, 21.6 g of
$\mathrm{FeCl_3\cdot6H_2O}$ and 11.2 g of $\mathrm{FeSO_4\cdot7H_2O}$ were mixed with 720 ml of ion-exchanged water (Milli-Q) in a plastic bottle (Thermo Scientific, 21040001) under ambient conditions. Co-precipitation was initialized by adding 80 ml of $\mathrm{NH_4OH}$, followed by vigorous mechanical stirring for 5 minutes. The cap was opened occasionally to release the excess $\mathrm{NH_3}^+$. Thereafter, 25 g of oleic acid was added to the bottle, the cap was closed, and the container was continuously shacked in vigorous vertical movements for 10 minutes, with occasional opening of the cap in between. The container was left open in a fume hood overnight to complete the reactions and to release the $\mathrm{NH_3}$. The NPs were purified the next day. First, the NPs were sedimented by placing the container on top of a cube-shaped NdFeB magnet (5 cm x 5 cm x 5 cm) for 10 min. The supernatant was removed and 400 ml of acetone was added to the mixture, followed by sonication at 37 kHz for 2 min to redisperse the sediment as aggregates. The NP aggregates were magnetically sedimented again within 2-3 min, resulting in a clear supernatant that was removed. Thereafter, 150 ml of toluene was added to re-disperse the NPs, and sonicated for 2 min. The NPs were precipitated by adding 200 ml acetone, leading the color of dispersion to change from brown to dark grayish. The NPs were again magnetically sedimented within 2 min, and the supernatant was removed. Finally, 50 ml of toluene was added, leading to spontaneous re-dispersion of the NPs by gently shaking the container. The NPs were then sonicated for 5 min with 37 kHz frequency to further assist the dispersion of NPs. The dispersion was left overnight in the fume hood to allow any remaining acetone to evaporate. The volume fraction of NPs in this stock dispersion was determined to be approximately 6.2 \% (Supplementary Note S1).

\justify \textbf{Characterisation of the iron oxide NPs}\\ 
\textit{Transmission electron microscopy (TEM):} NP morphology and size distribution were determined using a transmission electron microscope (JEOL JEM-2800, 200 kV). TEM sample was prepared by diluting the NP stock dispersion with toluene approximately in ratio 1:1000 and pipetting a droplet of the diluted dispersion on a TEM grid with a holey carbon film (Agar Scientific S147-4) and allowing the toluene to evaporate. From individual TEM images (Extended Data Fig. 1a), the NP size distribution (Extended Data Fig. 1b) was determined using image analysis software (ImageJ2/Fiji \cite{rueden2017imagej2, schindelin2012fiji}) by measuring the diameter of each particle manually along a randomly chosen axis. Approximately 350 particles were measured and the resulting histogram was fitted with a log-normal distribution, given by:
\begin{equation}
    y = y_0 + \frac{A}{x S \sqrt{2\pi}} \exp \left(-\frac{(ln(x) - M)^2}{2 S^2}\right) .
\end{equation}
The best fit was obtained with parameters given in Table 1, from which the mean ($\mu$) and standard deviation ($\sigma$) were obtained as
\begin{equation}
    \mu = e^{M+S^2/2} , \qquad \sigma^2 = e^{S^2+2M} \left(e^{S^2} -1 \right) 
\end{equation}
resulting in $\mu = 8.31 \pm 0.06 $ nm and $\sigma = 2.64 \pm 0.04 $ nm.
\begin{table}[H]
    \centering
\begin{tabular}{|c|c|c|c|}
\hline
$y_0$ & $A$ & $M$ & $S$\\
\hline
0 $\pm$ 1   & 751 $\pm$ 23 & 2.069 $\pm$ 0.007 & 0.310 $\pm$ 0.007 \\
\hline
\end{tabular}
\caption{Parameter of the log-normal fit of the NP distribution measured from TEM images.}
\end{table}

\justify \textit{X-ray diffraction (XRD):} Crystallographic structure of the iron oxide NPs was studied using X-ray diffraction (Rigaku SmartLab). Dry NP powder was prepared by allowing 500-1000 $\mathrm{\mu l}$ of stock NP dispersion to dry in a ceramic evaporating dish overnight. The dry powder was collected and deposited on a standard microscope glass slide. Large NP aggregates were crushed to a fine powder using another microscope slide. This fine powder on a glass slide was then used to measure XRD. The X-ray data was collected in the 2$\theta$ range of 5-90$\degree$ at a scanning speed of 1 degree/min with scan step of 0.04$\degree$  (Extended Data Fig. 1c).
\justify \textit{Raman spectroscopy:} Crystallographic structure of the iron oxide NPs were studied using Raman spectroscopy (Horiba LabRAM HR). Similar sample as for XRD was used. Measurements were carried out with 632.8 nm laser (HeNe laser) in the range of 50-1800 cm$^{-1}$. The laser beam was focused on the NP powder using a 50x objective lens. Data was collected for 120 seconds and averaged over two cycles. The laser power was limited to below 0.6 mW to avoid beam damage (Extended Data Fig. 1d).

\justify \textbf{Preparation of stock solution of AOT in dodecane}\\
A stock solution containing 300 mM of AOT in dodecane was prepared by mixing 4.01258 g of AOT and 19.7865 g of dodecane in a glass vial. The vial was sealed with parafilm to reduce moisture uptake from the ambient air. The vial was allowed to stand for two days to ensure complete dissolution of AOT. Solutions with lower AOT concentration (e.g. 150 mM) were prepared from the 300 mM stock solution by diluting with pure dodecane.

\justify \textbf{Preparation of electroferrofluids}\\
First, the stock dispersion of iron oxide NPs in toluene was sonicated (Fisherbrand FB11207) for 10 min at a 37 kHz and filtered through a polytetrafluoroethylene (PTFE) membrane with 0.45 ${\mathrm{\mu m}}$ pores (Thermo Scientific Titan3). 1 ml of the sonicated and filtered dispersion was mixed with 1 ml of a solution of AOT in dodecane with desired AOT concentration (e.g. 150 mM) in a glass vial. The vial was left open in a fume hood for ca. 36 hours until toluene was evaporated, which was confirmed with FTIR analysis (Nicolet 380, Thermo Fisher Scientific) by following the disappearance of the C-C vibration mode of the aromatic ring of toluene (Extended Data Fig. 1e). Finally, the moisture content was adjusted by placing the vial for ca. 15-30 hours in a desiccator cabinet (VWR 467-0130) at room temperature with 55\% relative humidity (created with a saturated aqueous solution of magnesium nitrate hexahydrate), resulting in the final electroferrofluid. The final electroferrofluid composition, i.e. volume fraction of NPs and AOT molarity, were estimated to be only slightly lower than in the stock NP dispersion and stock AOT solution (Supplementary Note S2). Throughout the text we refer to nominal AOT concentrations used in preparation of the electroferrofluid, which are close to the true values (Supplementary Note S3).

\justify \textbf{Magnetic characterization of the electroferrofluid (150 mM AOT)} Magnetic properties of the electroferrofluid (150 mM AOT) were measured with a vibrating sample magnetometer (QuantumDesign PPMS VSM). Sample was prepared by filling a 3 cm long capillary tube (0.22 mm inner diameter) with with ca. 0.5 ${\mathrm{\mu l}}$ of the electroferrofluid. The capillary was sealed with UV curable adhesive (Norland Optical Adhesive 61) and cured under UV lamp (Thorlabs Solis 365C). The paramagnetic background was removed by fitting the data at large fields and subtracting it from the measured data (Extended Data Fig. 1f). Magnetic susceptibility, saturation magnetization, and effective magnetic core diameter were obtained by fitting the measured magnetization loop with Langevin theory for superparamagnetism (Supplementary Note S3).

\justify \textbf{Filling microelectrode cells and making electrical connections}\\
Electrical connections to the microelectrode cells were done using thin insulated copper wires by attaching them to the cells with silver paste (SPI Supplies OK-SPI). The paste was allowed to dry at room temperature at least for two hours. The microelectrode cells were filled with the electroferrofluid using capillary action. The electroferrofluid was sonicated and filtered (similarly as during electroferrofluid preparation) just before filling the microelectrode cell for experiments on microscopic pattern formation (Fig. 2d,3,4,5 and Extended Data Fig. 5,7,8,9). In other experiments (Fig. 1e, 2b, Extended Data Fig. 4), the electroferrofluid was used without further sonication or filtering and small amounts of NP aggregates can be thus seen in the microscopy images.

\justify \textbf{COMSOL simulations}\\
Electric potential and electric field inside the microelectrode cells (Extended Data Fig. 2d,e) were simulated using finite element method (COMSOL Multiphysics 5.4). Electric Currents interface was used together with manufacturer specified cell geometry (Extended Data Fig. 2c) that were confirmed  (Extended Data Fig. 2b) using optical microscopy (Zeiss Z1 Imager and Zeiss Epiplan-Neofluar 50x/0.55). Following conductivity and relative permittivity values were used in the simulation: Glass: 1$\times 10^{-20}$ $\mathrm{S/m}$, 7.3; Sample space (dodecane): 1$\times10^{-8}$ $\mathrm{S/m}$, 2.0; ITO electrodes: 5$\times 10^{-6}$ $\mathrm{S/m}$, 3.6. 
 
\justify \textbf{Construction of the experimental setup for microscopic observation of the electroferrofluid under controlled electric and magnetic fields}
\justify \textit{Magnetic field:} A pair of small electromagnetic coils (GMW 11801523 and 11801524) connected to DC power supply (BK Precision 9205) was used to generate uniform magnetic fields (Extended Data Fig. 6a,b). The magnetic field between the coils was calibrated using a 3-axis teslameter (Senis 3MTS, Extended Data Fig. 6c-e). 
\justify \textit{Electric field:} A high resistance meter / electrometer (Keysight 2987A) was used to apply electric field across the sample and simultaneously measure the current, typically at 0.1 s intervals. Before each measurement, the electroferrofluid sample was ensured to be in the thermodynamic ground state by applying 0 V for 40 seconds, during which the electric current decayed close to 0 A and a homogeneous NP concentration was established.
\justify \textit{Microscopy:} The microelectode cell was illuminated in transmitted light configuration using an LED light source (Thorlabs MCWHLP1), a collimator (Thorlabs COP4-A) and a diffuser. Imaging was done with a 10x finite-conjugate objective lens (Nikon 10x/0.25 160/- WD5.6) or a 4x finite-conjugate objective lens (Nikon 4x/0.25 160/- WD25) connected to a 5 MP grayscale camera (Ximea MC050MG-SY). Image length scale was calibrated using a calibration target (Thorlabs R1L3S2P). Images were acquired using software provided by the camera manufacturer (Ximea xiCamTool 4.28). In most experiments, images were acquired at 100 frames per second (fps) with averaging of 5 consecutive frames to reduce noise, resulting in final acquisition rate of 20 fps. The Rosensweig-like instabilities (Fig. 4b,c) were imaged with 5 fps with averaging 5 consecutive images, leading to 1 fps, and the control experiments in polar solvent (Extended Data Fig. 3) at 30 fps with no image averaging. 

\justify \textbf{Control experiment with negatively charged superparamagnetic NPs in polar solvent (water)}\\
An aqueous ferrofluid consisting of maghemite NPs stabilized with citric acid (electrostatic double layer forces) was synthesized as described earlier \cite{Massart1981preparation}. A microelectrode cell was activated in a plasma oven (Henniker Plasma HPT-100) for 5 minutes to increase the hydrophilicity and to enable filling with the aqueous ferrofluid. The sample was imaged and electric and magnetic fields were applied with the same setup as in experiments with the electroferrofluid samples (Extended Data Fig. 3). 

\justify \textbf{I-V measurements}\\
I-V measurements (Fig. 2a) were carried out using the same setup used for the main experiments. A voltage staircase 0 $\rightarrow$ +3 $\rightarrow$ -3 $\rightarrow$ 0 V at 0.1 V intervals was applied and electric current was measured at each voltage after 30 second stabilization time. The measurement was performed three times in different microelectrode cells for 150 mM AOT in dodecane and the electroferrofluid and once for the iron oxide NPs in dodecane (0 mM AOT).

\justify \textbf{Image acquisition, processing and analysis}
\justify \textit{Image acquisition, bit depth conversion and contrast adjustment:} Microscopy images were acquired with 12-bit ADC and saved as 12/16 bit grayscale TIFF files. These high-bit images (raw data) were converted to 8-bit grayscale images with simultaneous histogram adjustment with ImageJ2/Fiji to improve contrast of the NP gradients and patterns (see Extended Data Fig. 7 for typical examples of raw and processed images).\cite{rueden2017imagej2, schindelin2012fiji} All shown microscopy images have been adjusted, unless otherwise stated in the text.
\justify \textit{Analysis of voltage-controlled magnetism:} The raw 12/16 bit images of the steady state ($U$ $>$ 0 V, Extended Data Fig. 5a, Fig. 2d) were divided pixel-by-pixel with the image of the thermodynamic ground state ($U$ = 0 V) with ImageJ/Fiji. This results in images consisting of real numbers close to 1.0 for each pixel (Extended Data Fig. 5b), that were saved as TIFF files with real (32-bit) values. These images were opened in MATLAB to calculate the natural logarithm (Extended Data Fig. 5c, Fig. 2e). From these images, the profiles (Fig. 2f) were obtained by averaging along the direction of the electrodes.
\justify \textit{Kymographic analysis:} Kymographs (Fig. 3e, 4e) were extracted using ImageJ2/Fiji by opening the image time series as an image stack, defining a 1-pixel wide region of interest (a line) along the electrodes and between a pair of a positive and a negative electrode, at a distance $d/4$ away from the edge of the positive electrode. Kymographs were then generated using the reslice command of ImageJ2/Fiji.\cite{rueden2017imagej2, schindelin2012fiji}
\justify \textit{Fourier analysis:} Fourier analysis of the transition from a steady-state gradient to a steady-state pattern (Extended Data Fig. 8e,9e) was carried out using a custom MATLAB code calculating the fast Fourier transform (FFT) of each row of pixels along the electrodes separately, followed by averaging the absolute values of the FFT over all pixel lanes of the image. The periodicity of patterns (Fig. 3d,4d) was obtained from the location of the peak in the Fourier transformation by applying a Gaussian fit. 

\justify \textbf{Analysis of dissipation}\\
The change in power dissipation upon pattern formation (Fig. 5d) was obtained as follows. First, the change in electric current upon application of magnetic field (resulting in pattern formation) was extracted from $I(t)$ by performing a baseline fitting and subtraction (Extended Data Fig. 10a-d). The steady-state change in the current ($\Delta I$ was then obtained after the change in current had stabilized (Extended Data Fig. 10e). Multiplication with the applied voltage led to the change in dissipation (Fig. 5d). This analysis also yields other useful information, such as the characteristic time scales of the growth and relaxation of the increase in dissipation (Extended data Fig. 10f).

\clearpage

\addcontentsline{toc}{section}{\numberline{}References}

\bibliographystyle{naturemag}
\bibliography{2.References}


\begin{flushleft}
\justify

\medskip
\justify
\textbf{Acknowledgements}
\justify
This work was carried out under the Academy of Finland’s Centres of Excellence Programme (2014-2019) and supported by ERC-2016-ADG-742829 DRIVEN. JVIT acknowledges funding from ERC (803937) and Academy of Finland (316219). Dr. Nikos Kyriakopoulos is acknowledged for assistance with FFT analysis and Dr. Mika Latikka for assistance with video processing. We acknowledge the facilities and technical support by Aalto University at OtaNano - Nanomicroscopy Centre (Aalto-NMC).
\medskip
\justify
\textbf{Author Contributions}
\justify
$^\dagger$These authors contributed equally. FS, TC and CR synthesised and characterised the electroferrofluids. CR designed and constructed the magnetic field setups assisted by TC. FS carried out the experiments with varying AOT concentrations and the I-V measurements. TC carried out the experiments on pattern formation and dissipation. TC, FS and CR analyzed the data and wrote the first version of the manuscript. CR compiled the figures and FS created the supplementary videos. OI supervised TC and contributed to the concepts. JVIT conceived the concepts and guided and supervised experimental work, data analysis and writing of the manuscript.

\bigskip

\newpage
\justify
\textbf{Additional information}

\smallskip
\justify
\textbf{Competing financial interests}
\justify
Authors have submitted a provisional patent application on electrically controllable ferrofluids.
\justify
\textbf{Correspondence and requests} 
\justify
All correspondence and requests for materials should be addressed to *jaakko.timonen@aalto.fi.

\end{flushleft}

\newpage

{\large\textbf{Extended Data}}

\renewcommand{\figurename}{Extended Data Figure}
\renewcommand{\thefigure}{\arabic{figure}}
\setcounter{figure}{0}

\begin{figure}[H]
\centering
\includegraphics[width=0.75\textwidth]{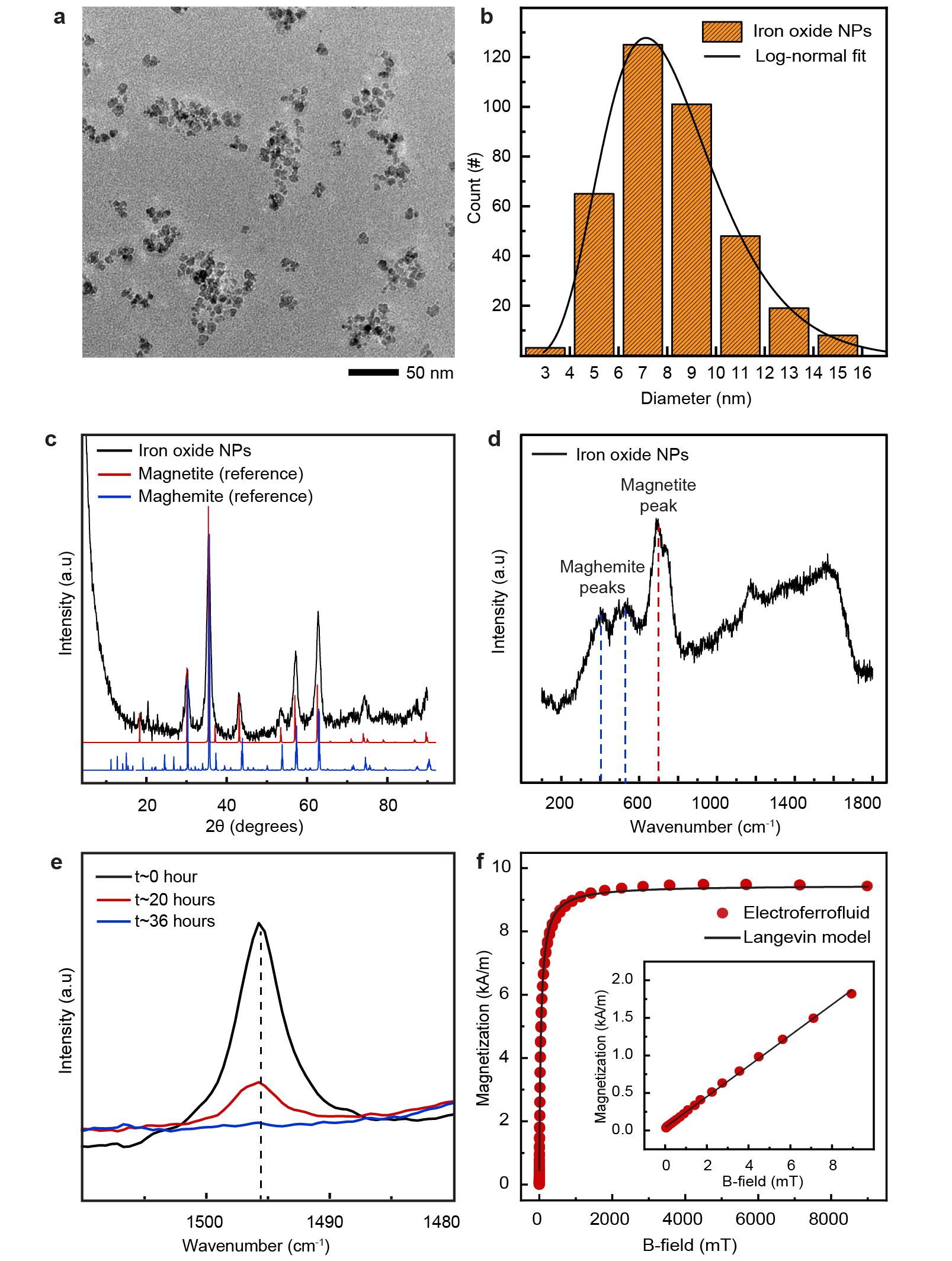}
\label{Extended_Data_Figure.01}
\caption {\textbf{Characterization of superparamagnetic NPs and the electroferrofluid.} \textbf{a,} A typical TEM image of the iron oxide NPs stabilized with oleic acid. 
\textbf{b,} NP diameter histogram obtained from TEM images with log-normal fit.
\textbf{c,} Powder XRD diffractogram of iron oxide NPs and reference diffractograms of magnetite and maghemite.\cite{okudera1996temperature,jorgensen2007formation}
\textbf{d,} Raman spectrum of iron oxide NPs and reference locations of the characteristic magnetite peak (ca. 660 ${\mathrm{cm^{-1}}}$) and maghemite peaks (300-500 ${\mathrm{cm^{-1}}}$).\cite{de1997raman} 
\textbf{e,} FTIR spectrum of mixture of iron oxide NPs in toluene and 150 mM AOT in dodecane for three data points during the evaporation of toluene at room temperature. The peak at 1496 ${\mathrm{cm^{-1}}}$ corresponds to the C-C vibration of aromatic ring of toluene.
\textbf{f,} Magnetization curve of the stock dispersion of iron oxide NPs in toluene and the best fit of Langevin model (See Supplementary Note S3 for model description and extracted parameters).
} 
\end{figure}

\begin{figure}[H]
\centering
\includegraphics[width=1\textwidth]{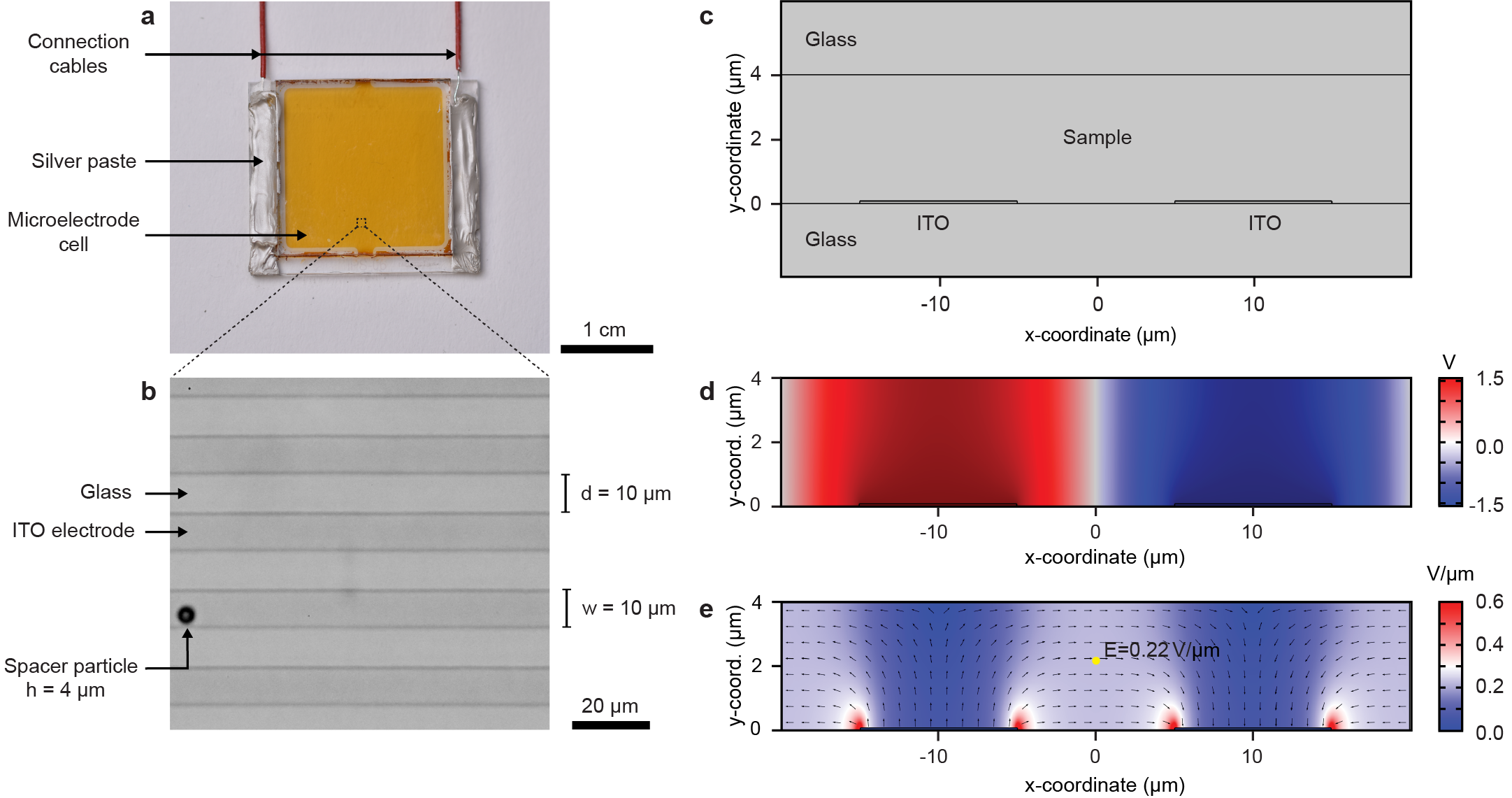}
\label{Extended_Data_Figure.02}
\caption{\textbf{Microelectrode cell geometry and the electric field within.} 
\textbf{a,} Photograph of a microelectrode cell filled with electroferrofluid. 
\textbf{b,} Micrograph of a microelectrode cell, showing lanes of interdigitated electrodes. 
\textbf{c,} Cross-sectional drawing of the microelectrode cell based on manufacturer specifications. 
\textbf{d,} Electric potential and \textbf{e,} electric field inside the microelectrode cell when a potential difference of 3 V is applied between neighbouring ITO electrodes with edge-to-edge distacen $d$ = 10 $\mu$m (COMSOL simulation). The electric field is approximately uniform between the electrodes (0.22 V$/\mu$m), with the exception of areas near the electrode corners where it is significantly stronger. In the main text, all quoted electric field strengths are nominal values calculated as $E$ = $U/d$ (0.30 V$/\mu$m in this COMSOL example), corresponding to the average field strength between two neighbouring electrodes on the bottom the microelectrode cell.
} 
\end{figure}

\begin{figure}[H]
\centering
\includegraphics[width=1\textwidth]{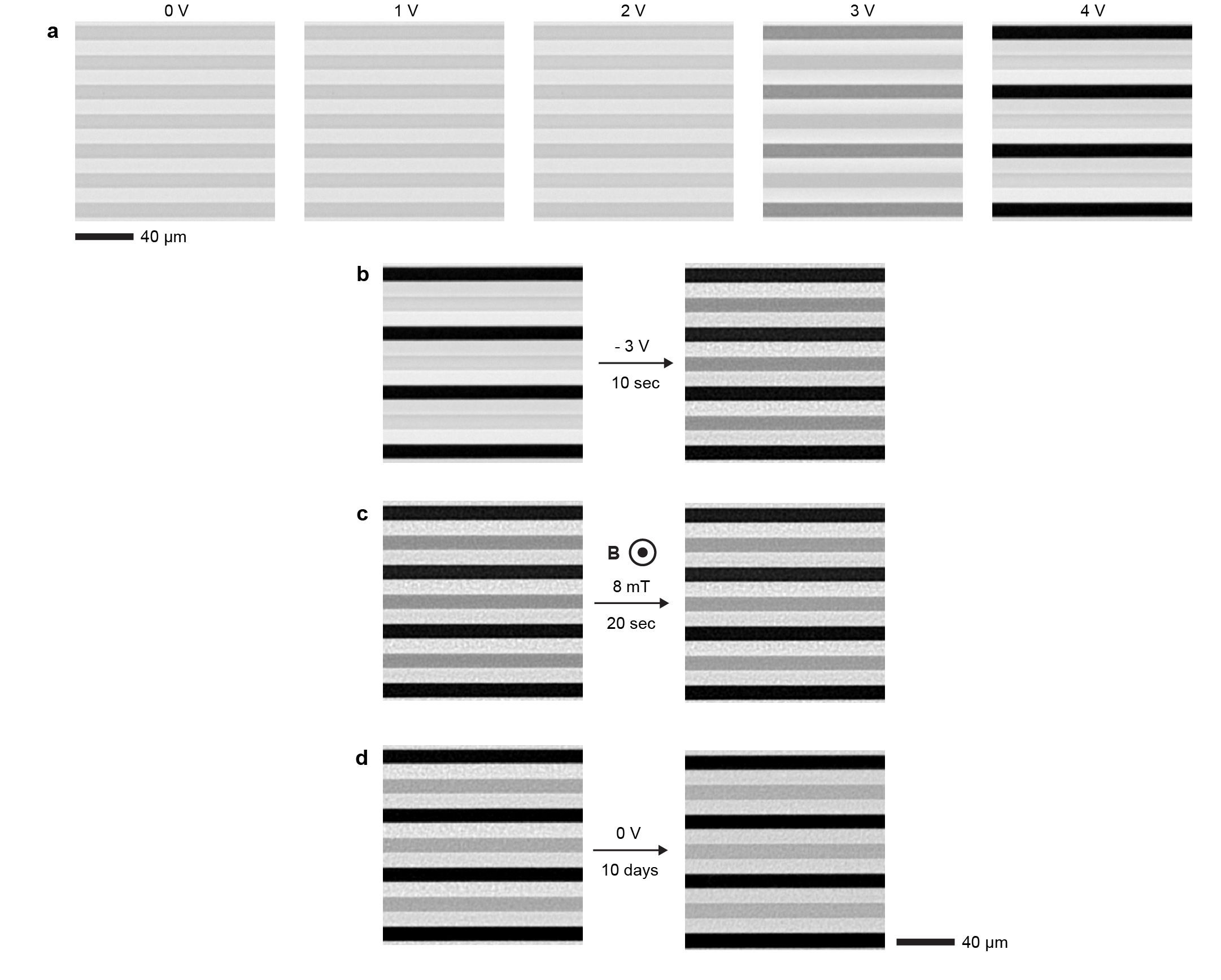}
\label{Extended_Data_Figure.03}
\caption{\textbf{Control experiment with negatively charged superparamagnetic NPs in polar solvent (water).} 
\textbf{a,} Microscopy images of a microelectrode cell filled with aqueous dispersion of negatively charged iron oxide NPs at different applied voltages. \textbf{b-d,} Microscopy images of the same microelectrode cell, showing no significant response of the sample to \textbf{b,} the change in polarity of the electric field, \textbf{c,} application of magnetic field, or \textbf{d,} long-term relaxation in absence of external electric and magnetic field.} 
\end{figure}

\begin{figure}[H]
\centering
\includegraphics[width=1\textwidth]{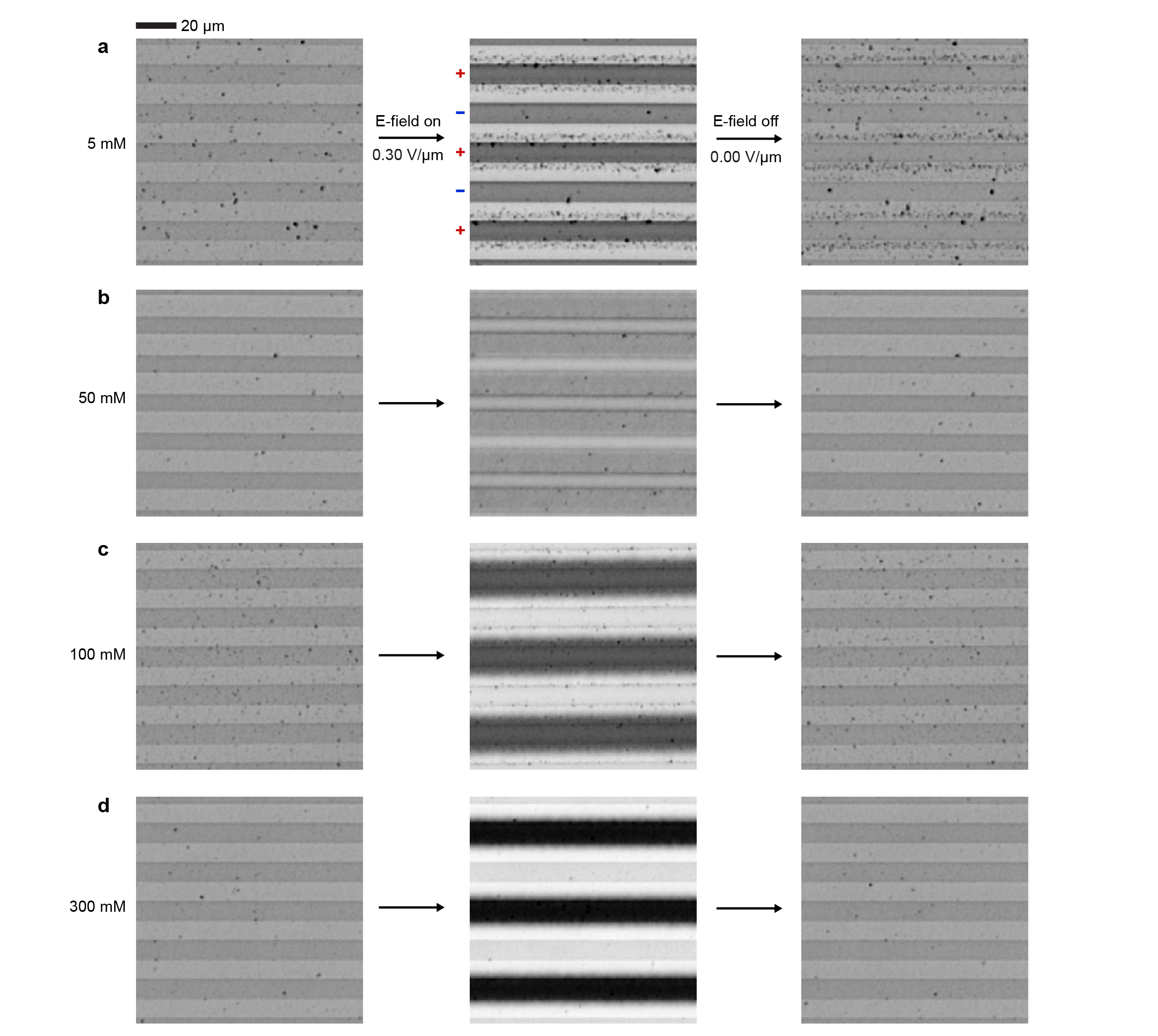}
\label{Extended_Data_Figure.04}
\caption{\textbf{Effect of charge control agent (AOT) concentration on electroferrofluid behavior.} Each row (a-d) contains three microscopy images of electroferrofluid of varying AOT concentration in a microelectrode cells before, during and after applying an electric field: 
\textbf{a,} 5 mM AOT: NPs form irreversible aggregates. 
\textbf{b,} 50 mM AOT: NPs form reversible weak concentration gradients. 
\textbf{c,} 100 mM AOT: NPs form reversible medium-strength concentration gradients. 
\textbf{d,} 300 mM AOT: NPs form reversible strong concentration gradients.}
\end{figure}

\begin{figure}[H]
\centering
\includegraphics[width=1\textwidth]{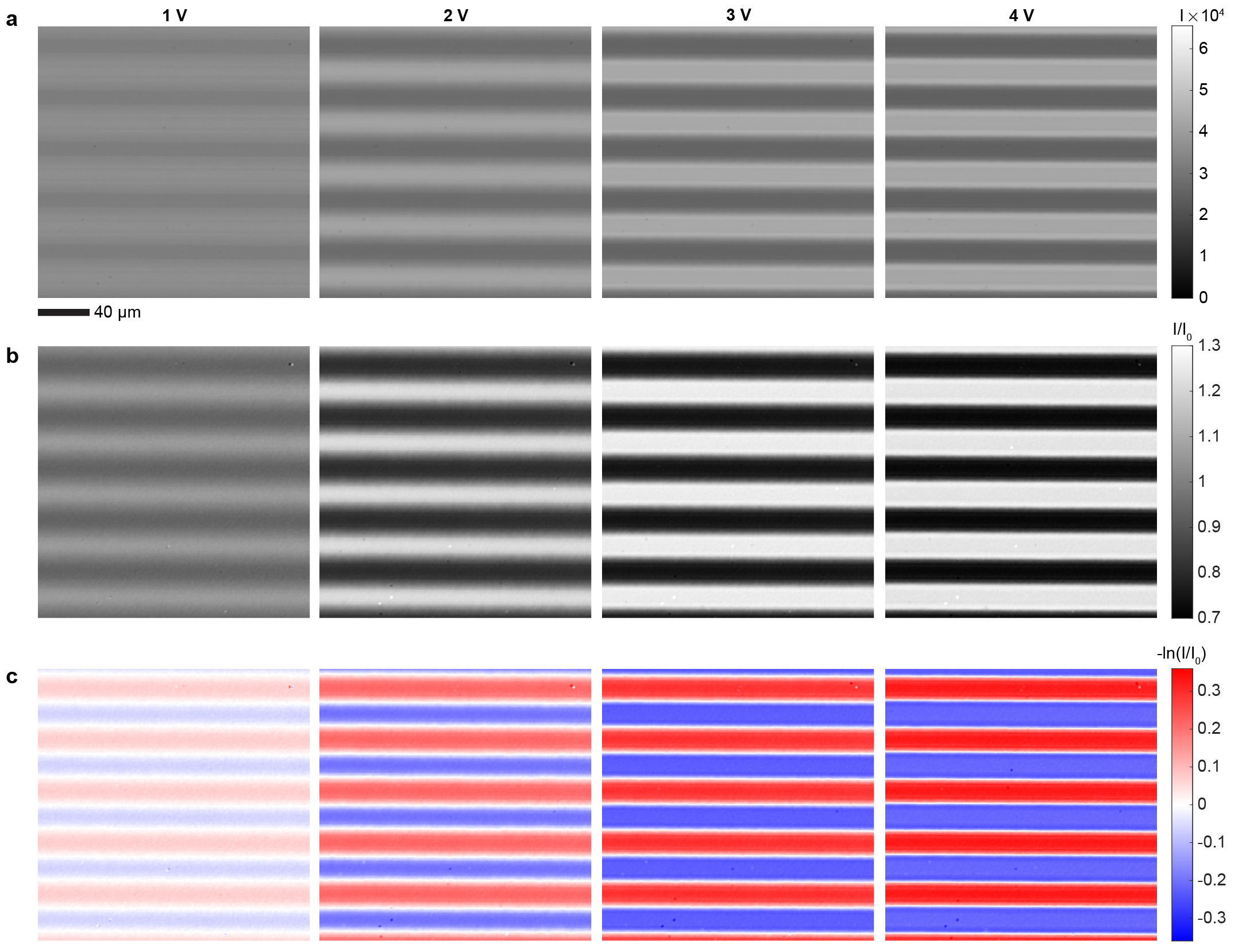}
\label{Extended_Data_Figure.05}
\caption{\textbf{Quantification of voltage-controlled magnetism in electroferrofluid.} 
\textbf{a,} Unprocessed microscopy images of the dissipative steady-states of the electroferrofluid (6 \% iron oxide NPs and 150 mM AOT) in a microelectrode cell at four different voltages. Each image is an intensity matrix $I$ with 16-bit intensity values assigned to each pixel.
\textbf{b,} Corresponding normalized images calculated by dividing each microscopy image pixel-by-pixel by a reference image $I_0$ (microscopy image of the equilibrium state, 0 V). 
\textbf{c,} Corresponding images showing $-\ln(I/I_0)$ that is approximately linearly proportional to change in NP concentration and magnetic response.} 
\end{figure}

\begin{figure}[H]
\centering
\includegraphics[width=1\textwidth]{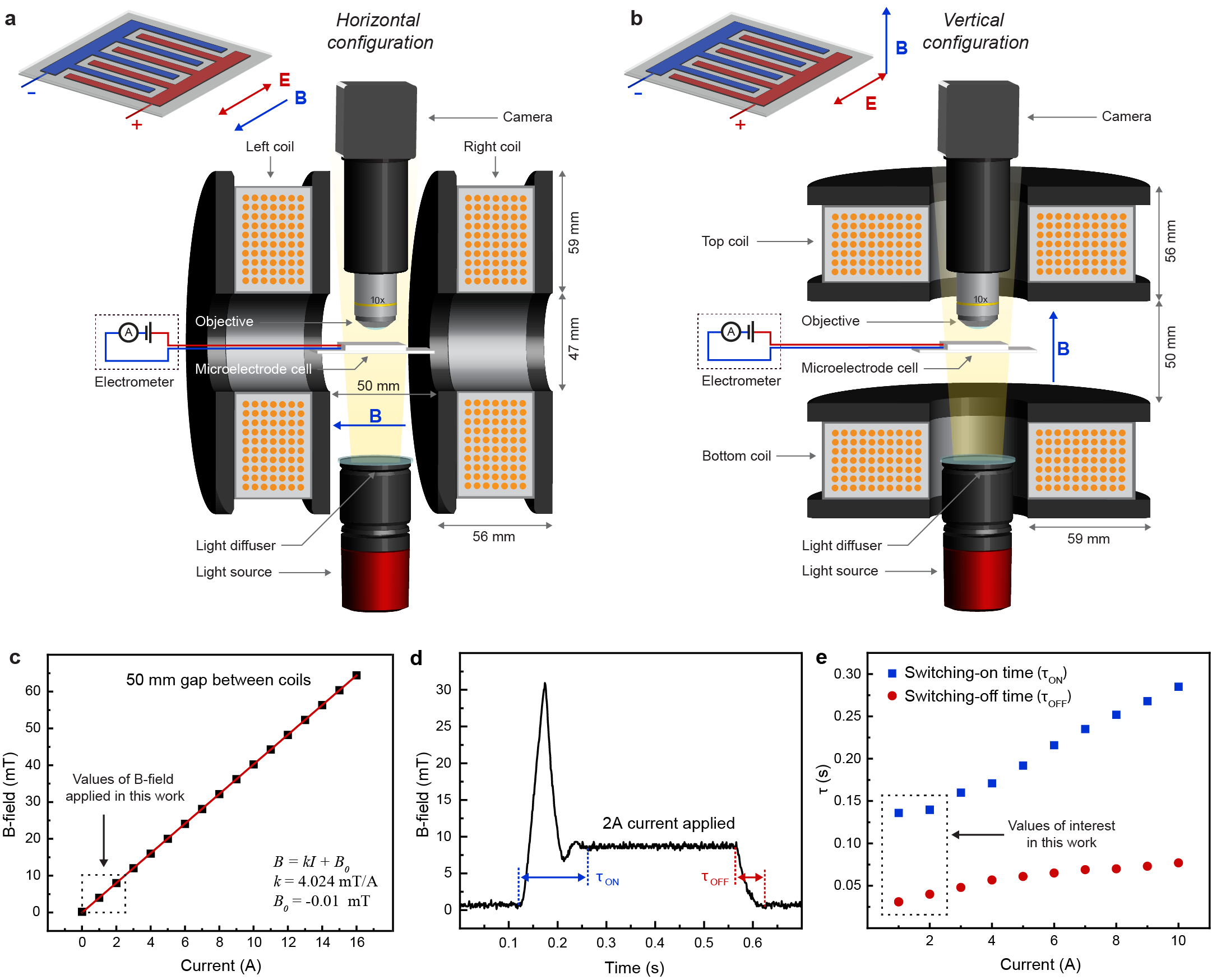}
\label{Extended_Data_Figure.06}
\caption{\textbf{Setup for simultaneous microscopic imaging and application of electric and magnetic fields.} 
\textbf{a,b} Drawings of the experimental setups used to produce \textbf{a,} horizontal (in-plane) and \textbf{b,} vertical (out-of-plane) magnetic fields.
\textbf{c,} Measured steady-state magnetic field as a function of applied current (applies to both in-plane and out-of-plane setups). 
\textbf{d,} Typical measured magnetic field stabilization behavior over time after turning on the electric current.
\textbf{e,} The time required to reach a steady-state magnetic field $\tau_\mathrm{{ON}}$ and the time required for the magnetic field to reduced to zero $\tau_\mathrm{{OFF}}$ upon turning on and off the electric current, respectively.} 
\end{figure}

\begin{figure}[H]
\centering
\includegraphics[width=1\textwidth]{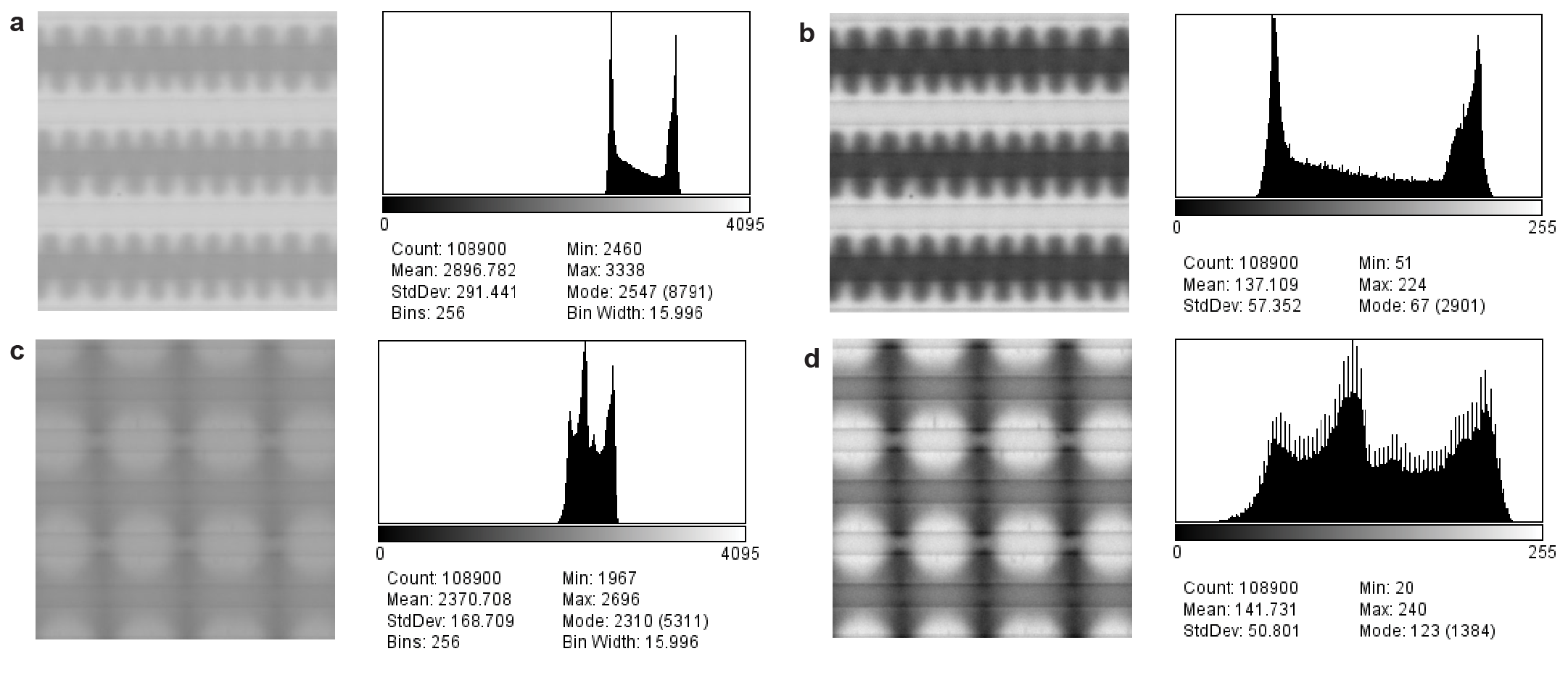}
\label{Extended_Data_Figure.07}
\caption{\textbf{Examples of image processing and contrast enhancement.} 
\textbf{a,} Unprocessed 12-bit image of a pattern in an out-of-plane magnetic field as obtained from the camera sensor, and the corresponding intensity histogram. All pixel intensity values are between 2460 and 3338. \textbf{b,} Processed 8-bit image and the corresponding histogram created by linear rescaling of the original image values between 2200 and 3500 to 8-bit range (0 and 255), thus maintaining all pixel values within the range of the new 8-bit image. \textbf{c,} Unprocessed 12-bit image of a pattern in an in-plane magnetic field and the corresponding histogram and \textbf{d,} corresponding processed 8-bit image and histogram.
}
\end{figure}

\begin{figure}[H]
\centering
\includegraphics[width=1\textwidth]{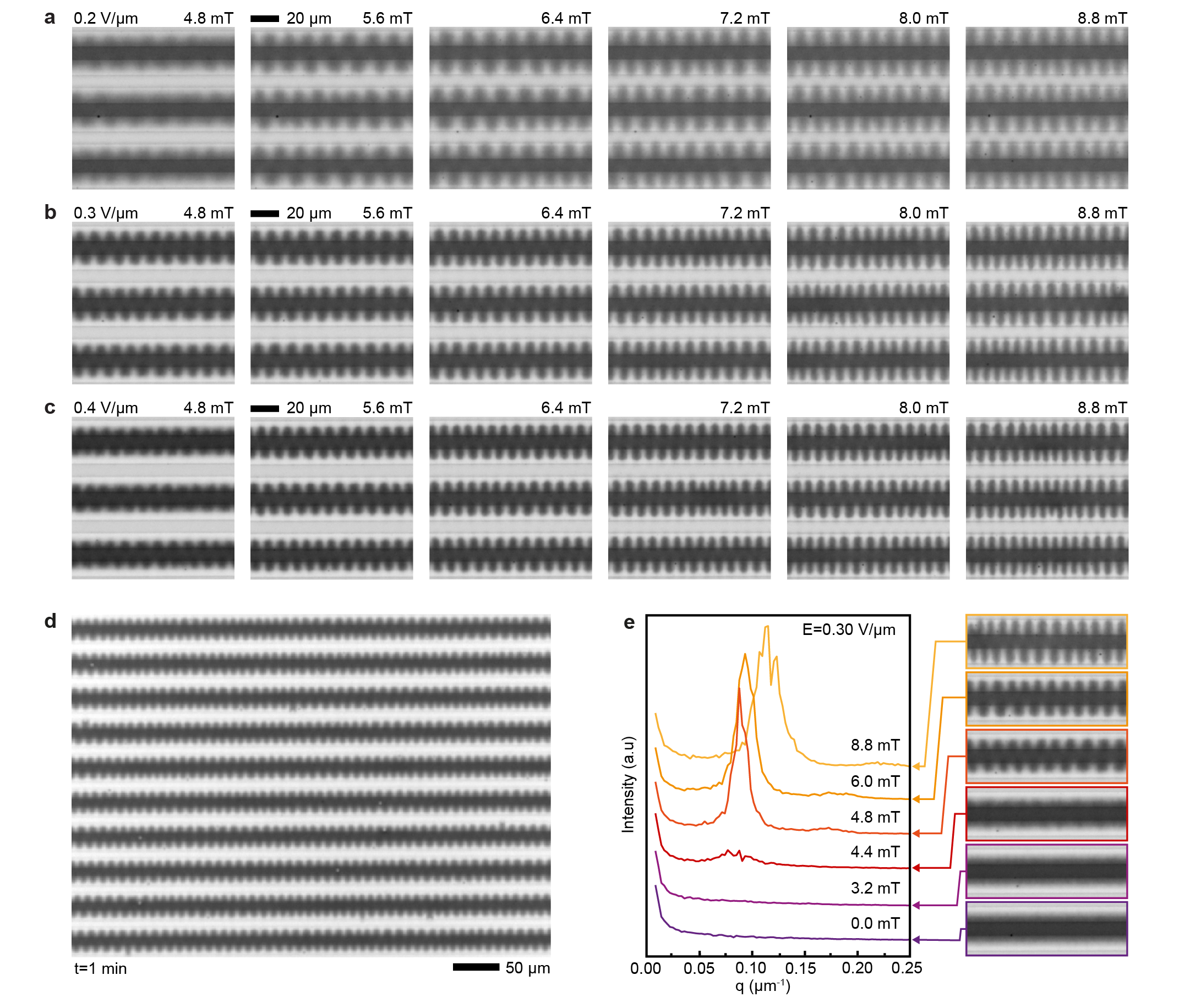}
\label{Extended_Data_Figure.08}
\caption{\textbf{Effect of electric and magnetic fields on pattern formation in the out-of-plane magnetic field.} 
\textbf{a-c,} Microscopy images of the dissipative steady-state of the electroferrofluid (6 \% iron oxide NPs and 150 mM AOT) in a microelectrode cell as a function of increasing magnetic field for \textbf{a,} $E$ = 0.2 V/${\mathrm{\mu m}}$, \textbf{b,} $E$ = 0.3 V/${\mathrm{\mu m}}$ and \textbf{c,} $E$ = 0.4 V/${\mathrm{\mu m}}$.
\textbf{d,} Low-magnification image of the steady-state pattern ($E$ = 0.3 V/${\mathrm{\mu m}}$, $B$ = 6.4 mT). 
\textbf{e,} 1D Fourier transformation of the pixel intensity profiles along the electrodes for different magnetic field strengths.
}
\end{figure}

\begin{figure}[H]
\centering
\includegraphics[width=1\textwidth]{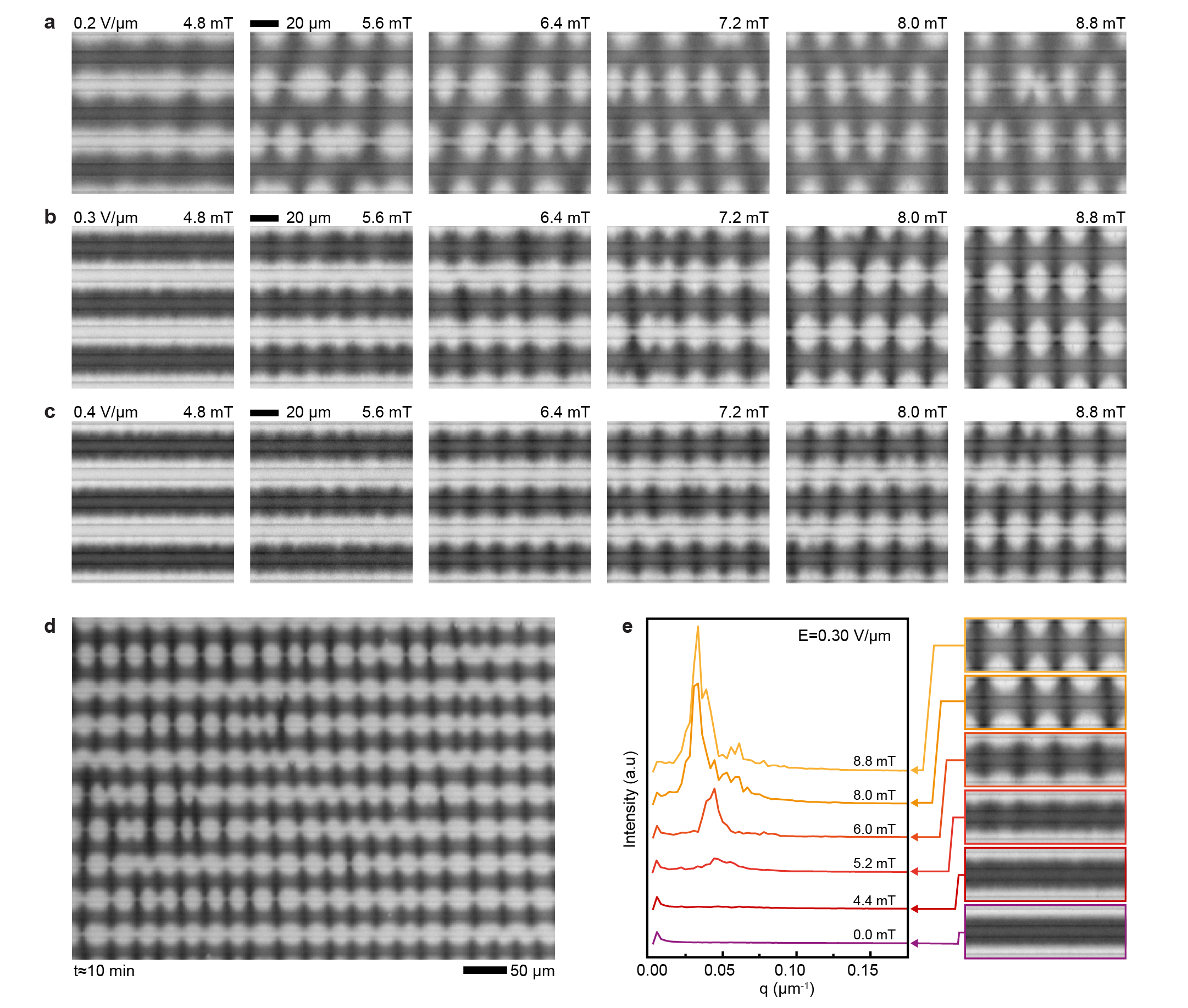}
\label{Extended_Data_Figure.09}
\caption{\textbf{Effect of electric and magnetic fields on pattern formation in in-plane magnetic field.} 
\textbf{a-c,} Microscopy images of the dissipative steady-state of the electroferrofluid (6 \% iron oxide NPs and 150 mM AOT) in a microelectrode cell as a function of increasing magnetic field for \textbf{a,} $E$ = 0.2 V/${\mathrm{\mu m}}$, \textbf{b,} $E$ = 0.3 V/${\mathrm{\mu m}}$ and \textbf{c,} $E$ = 0.4 V/${\mathrm{\mu m}}$.
\textbf{d,} Low-magnification image of the steady-state pattern ($E$ = 0.3 V/${\mathrm{\mu m}}$, $B$ = 8.0 mT). 
\textbf{e,} 1D Fourier transformations of the pixel intensity profiles along the electrodes for different magnetic field strengths.
}
\end{figure}

\begin{figure}[H]
\centering
\includegraphics[width=1\textwidth]{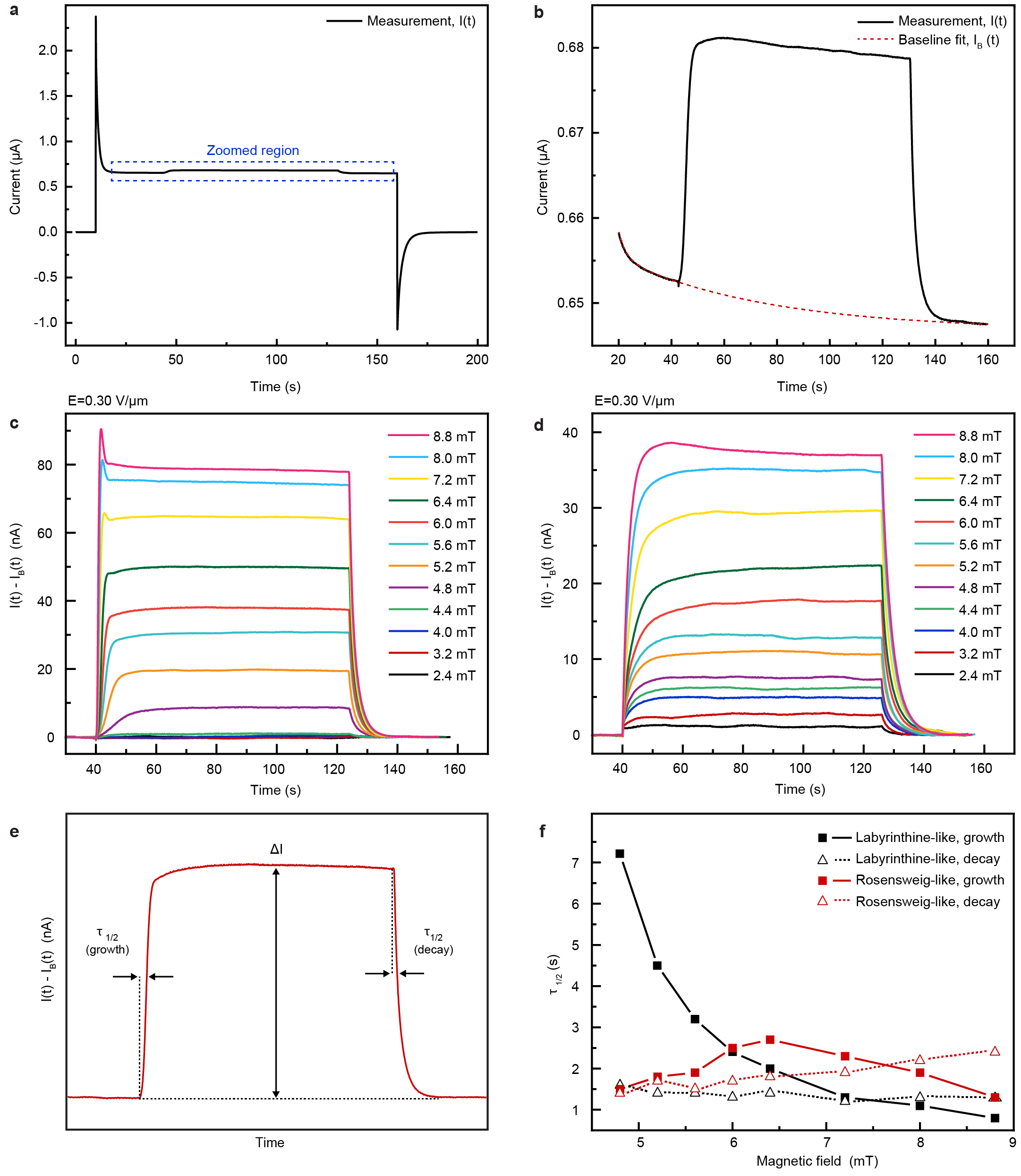}
\label{Extended_Data_Figure.10}
\caption{\textbf{Quantification of increase in dissipation during pattern formation.} 
\textbf{a,} Typical electric current as a function of time during formation of NP gradient and pattern in electric and magnetic fields and \textbf{b,} a close-up of the region where a change in current is seen during the pattern formation in magnetic field.
\textbf{c,} Change of current for various out-of-plane magnetic field strengths in $E$ = 0.3 V/${\mathrm{\mu m}}$. 
\textbf{d,} Change of current for various in-plane magnetic field strengths in $E$ = 0.3 V/${\mathrm{\mu m}}$.
\textbf{e,} Typical change of current with time constants $ \tau_{{1/2} ({\mathrm{growth}})}$ and $ \tau_{{1/2} ({\mathrm{decay}})}$ and steady-state change of current $\Delta I$ indicated. 
\textbf{f,} $ \tau_{{1/2} ({\mathrm{growth}})}$ and $ \tau_{{1/2} ({\mathrm{decay}})}$ as a function of magnetic field for labyrinthine-like and Rosensweig-like pattern formation.
} 
\end{figure}

\newpage

\renewcommand{\thetable}{S\arabic{table}}
\renewcommand{\theequation}{S\arabic{equation}}
\setcounter{table}{0}
\setcounter{equation}{0}

\newpage

\section*{Supplementary Materials}

\subsection*{List of supplementary videos}

\justify \textbf{Video 1. Formation of steady-state dissipative NP gradients in an electroferrofluid (6\% iron oxide NPs and 150 mM AOT) driven by electric field}\\
Video shows a series of microscopy images (video) and plots of electric current and electric field as a function of time in a microelectrode cell filled with an electroferrofluid (150 mM AOT) when a static electric field (0.3 V/${\mathrm{\mu m}}$) is applied for approximately 150 seconds. NPs form steady-state dissipative gradients in electric field that are reversible and disappear once the electric field is turned off. 

\justify \textbf{Video 2. Formation irreversible NP aggregates in an (electro)ferrofluid (6\% iron oxide NPs and 10 mM AOT) driven by electric field}\\
Video shows a series of microscopy images (video) and plots of electric current and electric field as a function of time in a microelectrode cell filled with an (electro)ferrofluid (10 mM AOT) when a static electric field (0.3 V/${\mathrm{\mu m}}$) is applied for approximately 150 seconds. NPs form aggregates in electric field that are irreversible and do not disappear once the electric field is turned off.

\justify \textbf{Video 3. Pattern formation in electroferrofluid (6\% iron oxide NPs and 150 mM AOT) in 6 mT out-of-plane magnetic field.}\\
Video shows a series of microscopy images (video) and plots of electric current, electric field and magnetic field as a function of time in a microelectrode cell filled with an electroferrofluid when a static electric field (0.3 V/${\mathrm{\mu m}}$) is applied and followed by application of magnetic field (6 mT, out-of-plane). NPs first form steady-state dissipative gradients in electric field and further a labyrinthine-like pattern in magnetic field. Both pattern and gradient are reversible once magnetic and electric fields are turned off, respectively.

\justify \textbf{Video 4. Pattern formation in electroferrofluid (6\% iron oxide NPs and 150 mM AOT) in increasing out-of-plane magnetic field.}\\
Video shows six series of microscopy images (video) of a microelectrode cell filled with an electroferrofluid when a static electric field (0.3 V/${\mathrm{\mu m}}$) is applied and followed by application of six different magnetic fields (from 3.2 to 8.8 mT, out-of-plane). In all cases, NPs first form steady-state dissipative gradients in electric field and further labyrinthine-like patterns in magnetic field with decreasing pattern periodicity with increasing magnetic field strength.

\justify \textbf{Video 5. Pattern formation in electroferrofluid (6\% iron oxide NPs and 150 mM AOT) in 8 mT in-plane magnetic field.} \\
Video shows a series of microscopy images (video) and plots of electric current, electric field and magnetic field as a function of time in a microelectrode cell filled with an electroferrofluid when a static electric field (0.3 V/${\mathrm{\mu m}}$) is applied and followed by application of magnetic field (8 mT, in-plane). NPs first form steady-state dissipative gradients in electric field and further a Rosensweig-like pattern in magnetic field. Both pattern and gradient are reversible once magnetic and electric fields are turned off, respectively.

\justify \textbf{Video 6. Pattern formation in electroferrofluid (6 \% iron oxide NPs and 150 mM AOT) in increasing in-plane magnetic field.}\\
Video shows six series of microscopy images (video) of a microelectrode cell filled with an electroferrofluid (150 mM AOT in dodecane) when a static electric field (0.3 V/${\mathrm{\mu m}}$) is applied and followed by application of six different magnetic fields (from 4.4 to 8.8 mT, in-plane). In all cases, NPs first form steady-state dissipative gradients in electric field and further Rosensweig-like patterns in magnetic field with increasing amplitude with increasing magnetic field strength.

\newpage
\setcounter{table}{0}
\setcounter{equation}{0}
\subsection*{Supplementary Note S1. Determination of the volumetric composition of the stock dispersion of iron oxide NPs in toluene}
Volume fraction of NPs in ferrofluid ($\Phi_{\mathrm{F}}$) was estimated by measuring and analyzing the mass of solid NPs ($m_{\mathrm{S}}$) in a known mass of the ferrofluid ($m_{\mathrm{F}}$) after evaporating the carrier solvent (toluene, with mass $m_{\mathrm{T}}$). Briefly, approximately $V_{\mathrm{F}}$ = 100 $\mathrm{\mu}$l of the ferrofluid was transferred into a 1.5 ml glass vial with known weight using a pipette (Eppendorf Z683809) and the mass of the vial with ferrofluid was measured, yielding the mass of the ferrofluid $m_{\mathrm{F}}$. The vial with ferrofluid was left open in a fume hood overnight to allow most of the toluene to evaporate. Remaining toluene (and other volatiles that we assumed they have negligible amounts) were removed in oven (Memmert UF30) at 70 $\degree$ for 30 hours. The remaining mass determined with analytical balance was $m_{\mathrm{S}}$, thus yielding mass of evaporated toluene as $m_{\mathrm{T}}$ = $m_{\mathrm{F}}-m_{\mathrm{S}}$. The volume fraction of solid content in the toluene ferrofluid is obtained as:
\begin{equation}
\Phi_{\mathrm{F}} = 
\frac{V_{\mathrm{S}}}{V_{\mathrm{F}}} =
\frac{V_{\mathrm{F}}-V_{\mathrm{T}}}{V_{\mathrm{F}}} =
1-\frac{V_{\mathrm{T}}}{V_{\mathrm{F}}} =
1-\frac{m_{\mathrm{T}}}{\rho_{\mathrm{T}}} \frac{\rho_{\mathrm{F}}}{m_{\mathrm{F}}},
\end{equation}
where $\rho_{\mathrm{F}}$ is the density of the ferrofluid and $\rho_{\mathrm{T}}$ density of toluene. We obtained approximate values for these by measuring the mass of a known volume of the liquids using an analytical balance and a positive displacement pipette (Eppendorf multipette E3x): $\rho_{\mathrm{F}}$ = 1.108 $\pm$ 0.007 g/ml is and $\rho_{\mathrm{T}}$ = 0.854 $\pm$ 0.001 g/ml. The volume fraction measurement was performed for three samples (Table S1). The average volume fraction of NPs calculated from the three measurements is 6.2 \% and the volume fraction of toluene is thus 93.8 \%. We note that in this approach the free surfactants (non-volatile) are also included in the volume fraction of NPs. We also note the relative error of ca. 1.3 \% when comparing measured toluene density and literature value (0.865 g/ml), and used the measured value that we assume to have similar systematic error as the ferrofluid density. 

\begin{table}[H]
	\centering
	\begin{tabular}{ |c|c|c|c|c| } 
		\hline
		Sample number & \(m_{\mathrm{F}}\) (mg) & \(m_{\mathrm{S}}\) (mg) & \(m_{\mathrm{T}}\) (mg) & \(\Phi_{\mathrm{F}}\) (\%) \\ 
		\hline
		1 & 101.2 & 28.0 & 73.2 & 0.061 \\
		2 & 100.9 & 28.0 & 72.9 & 0.063 \\
		3 & 100.6 & 27.8 & 72.8 & 0.061 \\
		\hline
	\end{tabular}
	\caption{Data used to calculate the volume fraction of the solid content in the ferrofluid in toluene}
\end{table}

\subsection*{Supplementary Note S2. Determination of the volumetric composition of the electroferrofluid (150 mM AOT)}
Standard electroferrofluid with 150 mM AOT was prepared by mixing $V_{\mathrm{F}}$ = 100 $\mathrm{\mu l}$ of ferrofluid (Supplementary Note S1) and $V_{\mathrm{AOT/DD}}$ = 100 $\mathrm{\mu l}$ of 150 mM AOT in dodecane, and allowing toluene to evaporate completely at room temperature, leading to final electroferrofluid with volume $V_{\mathrm{EF}}$. Assuming no loss of dodecane or significant moisture uptake during the evaporation, the volume fraction of NPs ($\Phi_{\mathrm{NPs}}$) and the volume fraction of AOT ($\Phi_{\mathrm{AOT}}$) in the final electroferrofluid can be calculated as
\begin{equation}
\Phi_{\mathrm{NPs}} = \frac{V_{\mathrm{NPs}}}{V_{\mathrm{EF}}} = 
\frac{V_{\mathrm{F}} \Phi_{\mathrm{F}}}{V_{\mathrm{AOT/DD}} + V_{\mathrm{F}} \Phi_{\mathrm{F}}},
\end{equation}
and
\begin{equation}
\Phi_{\mathrm{AOT}} = \frac{V_{\mathrm{AOT}}}{V_{\mathrm{EF}}} = 
\frac{V_{\mathrm{AOT/DD}} \Phi_{\mathrm{AOT/DD}}}{V_{\mathrm{AOT/DD}} + V_{\mathrm{F}} \Phi_{\mathrm{F}}}. 
\end{equation}
Inserting the above values to the equations together with $\Phi_{\mathrm{F}}=0.062$ from Supplementary Note S1 and the volume fraction of AOT in 150 mM AOT in dodecane, that was calculated as
$\Phi_{\mathrm{AOT/DD}}$ = (0.15 mol/l $\cdot$ 444.5583 g/mol) / 1100 g/l $\approx 0.061$, we obtain $\Phi_{\mathrm{NPs}} \approx 5.8$ \% and $\Phi_{\mathrm{AOT}} \approx 5.7$ \%. Due to the several approximations and experimental uncertainties, we conclude that the final electroferrofluid with 150 mM AOT contains (volumetrically) approximately 6 \% of iron oxide NPs (including oleic acid stabilizer), 6 \% of charge control agent AOT and 88 \% of dodecane. In addition, there is a small amount of water in the electroferrofluid that originates from preparation of the electroferrofluid under ambient conditions and incubation in the humidity chamber. The volume fraction of water was estimated to be approximately 0.5 \% to 1.0 \% using Karl Fischer titration (Mettler Toledo Coulometric KF Titrator C30S).

\subsection*{Supplementary Note S3. Magnetic properties of the electroferrofluid}
The data collected with the magnetometer as described in the Methods were then analysed as follows. The paramagnetic background from sample holder was subtracted from the measured data performing a linear fit of the data at the highest field. Positive half of the magnetic loop of the electroferrofluid is shown in Extended data Fig. 1f. The data were interpolated with the Langevin function as:

\begin{equation} \label{eu_eqn1}
M = nm \left[ \coth \left( \frac{mB}{{k_\mathrm{B}T}}\right) - \left(\frac{mB}{{k_\mathrm{B}T}}\right)^{-1} \right],
\end{equation}
where $M$ is the magnetization of the sample, $B$ is the applied magnetic field, $k_\mathrm{B}$ is the Boltzmann constant, $T$ is the temperature, $n$ is the concentration of the particles (measured in $\mathrm{m^{-3}}$), and $m$ is the average NP moment (measured in $\mathrm{A m^2}$). The results of the interpolation are summarized in Table S2. 

\begin{table}[h!]
	\centering
	\begin{tabular}{ |c|c| } 
		\hline
		$m_{\mathrm{EF}}$ ($\mathrm{Am^2}$) & $n_{\mathrm{EF}}$ ($\mathrm{m^{-3}}$) \\ 
		\hline
		$(2.03 \pm 0.08)\times 10^{-19}$  & $(4.4 \pm 0.2)\times 10^{22}$  \\
		\hline
	\end{tabular}
	\caption{Results for NP moment ($m$) and concentration of the NPs ($n$) for the electroferrofluid (EF)}
\end{table}
Comparing the magnetic moment of the NPs with the magnetic moment of the bulk magnetic material ($M_{\mathrm{b}}$) can allow us to derive different useful quantities. After simple calculations from the Langevin equation we can write that: 
\begin{equation} \label{eu_eqn}
d=\left(\frac{6 m}{\pi M_{\mathrm{b}} }\right)^{\frac{1}{3}} ,\qquad \phi = V n = \frac{\pi}{6}d^3n , \qquad M_s = \phi M_{\mathrm{b}} = n m , \qquad \chi = \frac{\mu_0 m M_{\mathrm{s}}}{3 k_\mathrm{B} T},
\end{equation}
where $d$ is the magnetic core diameter, $\phi$ is the volume fraction of the magnetic material (iron oxide) in the electroferrofluid, $M_{\mathrm{b}}$ $\approx 0.56$ T is saturation magnetization of magnetite \cite{Buschow2006Handbook}, $M_\mathrm{s}$ is the saturation magnetization of the fluid and $\chi$ is its magnetic susceptibility.

\begin{table}[ht!]
	\centering
	\begin{tabular}{ |c|c|c|c| }
		\hline
		$d_{\mathrm{EF}}$ (nm) & $\phi_{\mathrm{EF}}$ & $M_{\mathrm{S/EF}}$ (kA/m) & $\chi_{\mathrm{EF}}$ \\ 
		\hline
		9.5 $\pm$ 0.4 & 2.0 $\pm$ 0.1  & 9.01 $\pm$ 0.05  & 0.19 $\pm$ 0.01  \\
		\hline
	\end{tabular}
	\caption{Results of the calculation with formulas S5 for the electroferrofluid (EF)}
\end{table}

One can immediately notice  that the volume fraction of the magnetic material $\phi_{\mathrm{EF}}$ is lower than the volume fraction of the solid content $\Phi_{\mathrm{EF}}$ calculated previously in Supplementary Notes S1 and S3. This is due to the fact that volume fraction $\phi_{\mathrm{EF}}$ calculated here corresponds only to the magnetic core of NPs in electroferrofluid; whereas, the non-magnetic oleic acid shell that encapsulates the magnetic core also contributes to solid content of NPs in calculations in Supplementary Notes S1 and S3. 


\end{document}